\documentclass[aps,prl,floats,twocolumn,showpacs,superscriptaddress]{revtex4}

\usepackage{natbib}
\usepackage[utf8]{inputenc}

\usepackage{amsmath, amsthm, amssymb}
\usepackage{enumitem}
\usepackage{graphics,graphicx}% Include figure files
\usepackage{epsfig}
\usepackage{times}% Times fonts
\usepackage{dcolumn}% Align table columns on decimal point
\usepackage{url}
\usepackage{color}

\usepackage[all]{xy}
\usepackage{mdwlist}

\begin{document}

\renewcommand*\thesection{\arabic{section}}
\newcommand{\beq}{\begin{equation}}
\newcommand{\eeq}{\end{equation}}
\newcommand{\sss}{\scriptscriptstyle}
\newcommand{\rev}[1]{{\color{black} #1}}
\newcommand{\rerev}[1]{{\color{black} #1}}
\newcommand{\average}[1]{{\left\langle {#1} \right\rangle}}

\title{On the dynamical interplay between awareness and epidemic spreading in multiplex networks}

\author{Clara Granell}
\affiliation{Departament d'Enginyeria Inform\`atica i Matem\`atiques,
Universitat Rovira i Virgili, 43007 Tarragona, Spain}

\author{Sergio G\'omez}
\affiliation{Departament d'Enginyeria Inform\`atica i Matem\`atiques,
Universitat Rovira i Virgili, 43007 Tarragona, Spain}

\author{Alex Arenas}
%\email{alexandre.arenas@urv.cat}
\affiliation{Departament d'Enginyeria Inform\`atica i Matem\`atiques,
Universitat Rovira i Virgili, 43007 Tarragona, Spain}
\affiliation{IPHES, Institut Catal\`a de Paleoecologia Humana i Evoluci\'o Social, C/Escorxador s/n, 43003 Tarragona, Spain}

\begin{abstract}
We present the analysis of the interrelation between two processes accounting for the spreading of an epidemics, and the information awareness to prevent its infection, on top of multiplex networks. This scenario is representative of an epidemic process spreading on a network of persistent real contacts, and a cyclic information awareness process diffusing in the network of virtual social contacts between the same individuals. The topology corresponds to a multiplex network where two diffusive processes are interacting affecting each other. The analysis using a Microscopic Markov Chain Approach (MMCA) reveals the phase diagram of the incidence of the epidemics and allows to capture the evolution of the epidemic threshold depending on the topological structure of the multiplex and the interrelation with the awareness process. Interestingly, the critical point for the onset of the epidemics has a critical value (meta-critical point) defined by the awareness dynamics and the topology of the virtual network, from which the onset increases and the epidemics incidence decreases.
\end{abstract}


\maketitle

Real complex systems are often composed of several layers of networks interrelated with each other; when the actors in these different layers of networks are the same we call them multiplex networks. The understanding of the emergent physical phenomena on multiplex networks is gaining much attention \cite{kurant06,mucha10,Szell10,gardenes12,baxter12,cozzo12,bianconi13,gomez13} as a particular case of interdependent networks \cite{buldyrev10,gao11}. In particular, multiplex networks represent the natural way to describe social interactions that occur at different contexts or in different categories. For example, people have a series of persistent contacts in the day life with family, friends and coworkers that form the network of physical contacts, while at the same time, the same actors are connected using online social networks with the previous mentioned contacts and also probably with others. These different layers can support different dynamical processes, e.g.\ in online social networks actors exchange information in any form, while in the physical network actors exchange also biological elements that can carry on diseases.

The described scenario is a good proxy to analyze the interplay between information spreading of the awareness to a certain epidemics, and the epidemic infection itself, in a certain networked population. The importance of understanding this interplay relies on the consequences the awareness can have on the outbreak of the epidemics and  its incidence. Several works addressed the problem from different perspectives \cite{bagnoli07,funk09,funk11,marceau11,hatzopoulos11,xu12} considering, for example, the risk perception, behavioral changes, or competing viral agents.

\begin{figure}[tbp]
\begin{center}
  \includegraphics[width=0.85\columnwidth,clip=0]{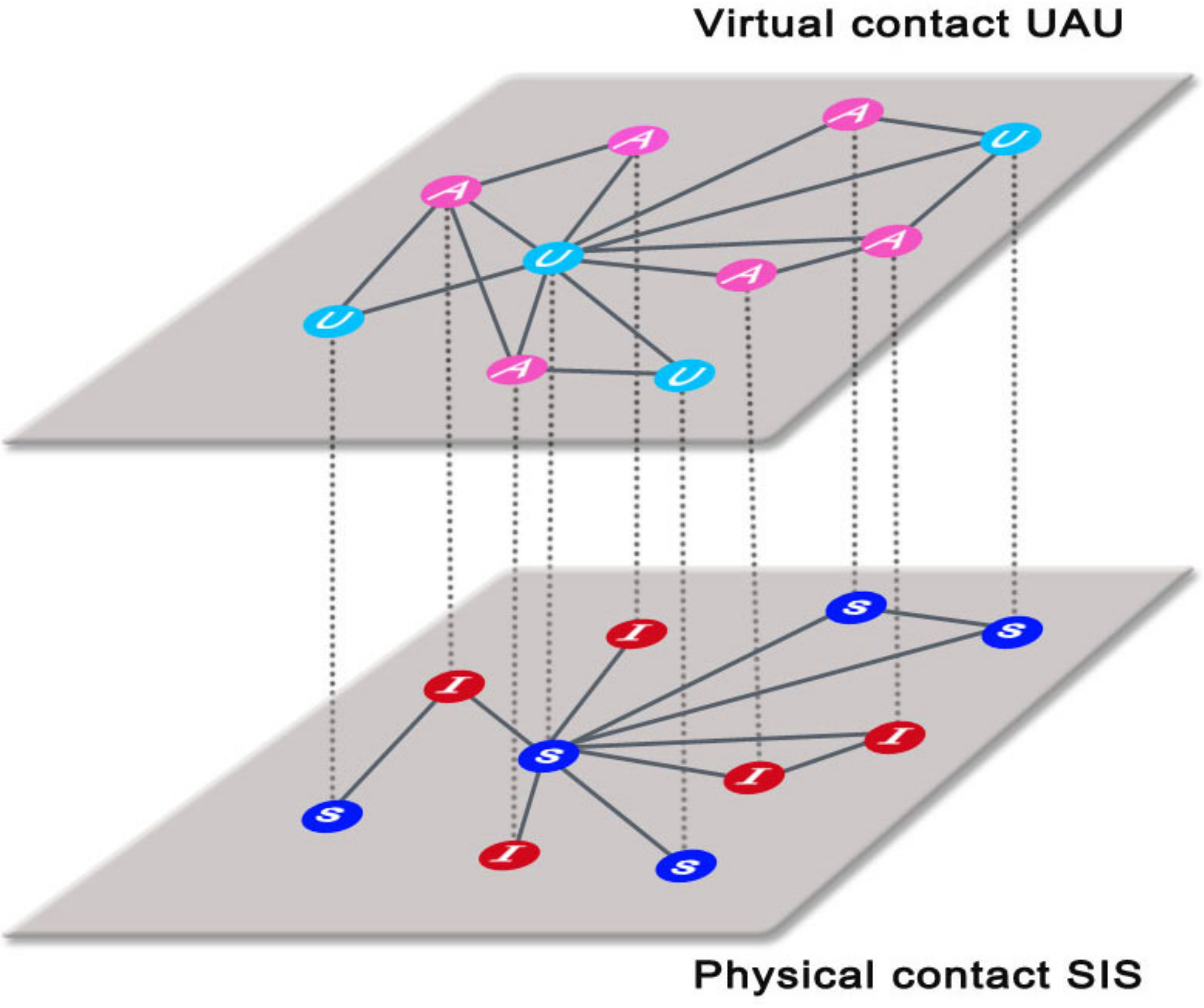}
\caption{(color online) Sketch of the multiplex structure type used in this work. The upper layer (virtual contact) is supporting the spreading of awareness, nodes have two possible states: unaware (U) or aware (A). The lower layer (physical contact) corresponds to the network where the epidemic spreading takes place. The nodes are the same actors than in the upper layer, but here their state can be: susceptible (S) or infected (I).}
\label{fig1}
\end{center}
\end{figure}

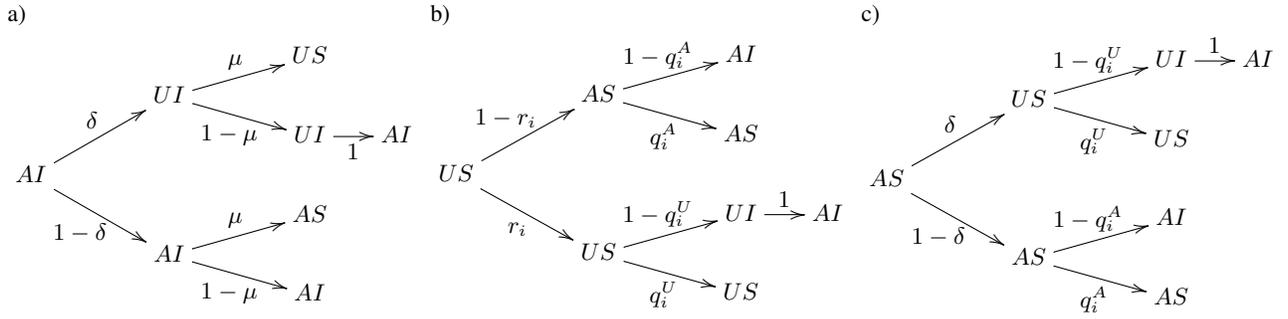
\begin{figure*}[tbp]

\begin{center}
\begin{tabular}{lll}
  a) & b) & c)
\\

$
\xymatrixcolsep{5mm}
\xymatrixrowsep{1mm}
\vcenter{\xymatrix{
  && && US
  \\
  && UI \ar[urr]^{\mbox{$\mu$}} \ar[drr]_{\mbox{$1-\mu$}}
  \\
  && && UI \ar[r]_{\mbox{$1$}} & AI
  \\
  AI \ar[uurr]^{\mbox{$\delta$}} \ar[ddrr]_{\mbox{$1-\delta$}}
  \\
  && && AS
  \\
  && AI \ar[urr]^{\mbox{$\mu$}} \ar[drr]_{\mbox{$1-\mu$}}
  \\
  && && AI
}}
$

&

$
\xymatrixcolsep{5mm}
\xymatrixrowsep{1mm}
\vcenter{\xymatrix{
  && && AI
  \\
  && AS \ar[urr]^{\mbox{$1-q_i^A$}} \ar[drr]_{\mbox{$q_i^A$}}
  \\
  && && AS
  \\
  US \ar[uurr]^{\mbox{$1-r_i$}} \ar[ddrr]_{\mbox{$r_i$}}
  \\
  && && UI \ar[r]^{\mbox{$1$}} & AI
  \\
  && US \ar[urr]^{\mbox{$1-q_i^U$}} \ar[drr]_{\mbox{$q_i^U$}}
  \\
  && && US
}}
$

&

$
\xymatrixcolsep{5mm}
\xymatrixrowsep{1mm}
\vcenter{\xymatrix{
  && && UI \ar[r]^{\mbox{$1$}} & AI
  \\
  && US \ar[urr]^{\mbox{$1-q_i^U$}} \ar[drr]_{\mbox{$q_i^U$}}
  \\
  && && US
  \\
  AS \ar[uurr]^{\mbox{$\delta$}} \ar[ddrr]_{\mbox{$1-\delta$}}
  \\
  && && AI
  \\
  && AS \ar[urr]^{\mbox{$1-q_i^A$}} \ar[drr]_{\mbox{$q_i^A$}}
  \\
  && && AS
}}
$
\end{tabular}
\end{center}

\caption{Transition probability trees for the states of the UAU-SIS dynamics in the multiplex per time step.
%A time step consists of an update of the states in the virtual network, and a consecutive update in the physical network.
The notation is: (AI) aware-infected, (AS) aware-susceptible, (UI) unaware-infected, (US) unaware-susceptible, $\delta$ transition probability from aware to unaware, $\mu$ transition probability from infected to susceptible, $r_i$ transition probability from unaware to aware given by neighbors, $q_i^{A}$ transition probability from susceptible to infected, if node is aware, given by neighbors, and $q_i^{\mbox \scriptsize U}$ transition probability from susceptible to infected, if node is unaware, given by neighbors.
}
\label{fig2}
\end{figure*}

In this letter, we propose the use of the Microscopic Markov Chain Approach (MMCA) \cite{chakrabarti08,gomez10,gomez11} to understand the interplay between an epidemic spreading process, and a cyclic spreading of awareness process in quenched multiplex networks. The multiplex corresponds to a two layer network, one where the dynamics of the awareness evolves and another where the epidemic process spreads. The approximation using MMCA has an accuracy up to $2.5\%$ error for the prediction of the epidemic threshold and epidemic incidence. The error has been computed comparing with extensive Monte Carlo simulations of the same system.
\rev {This setup is an abstraction for those epidemics that satisfy the dynamics of susceptible-infected-susceptible processes (SIS) coexisting with a cyclic process of awareness spreading satisfying the cycle unaware-aware-unaware (UAU). It could represent the interrelated dynamics of those epidemics like influenza, with a marked seasonal character, and the word of mouth of aware individuals advising their social acquaintances to take a flu-shot. Note that here we are not considering the effect of the media on the vaccination campaign.}

Let us start by defining the specific setup we analyze. We use a multiplex, see Fig.~\ref{fig1}, with different connectivity at each layer, corresponding to the layer of {\em physical} persistent social contacts (those that can infect you), and to the layer of {\em virtual} contacts (those that communicate with you but are not necessarily in physical contact, e.g.\ Facebook friends, etc.). Note that we are not using a framework of general interdependent networks, because the actors in both layers are the same. However, as observed in interdependent networks \cite{saumell12} the interrelation between two different structures is responsible for the emergence of new physical effects on the epidemic onset and prevalence of the epidemics.

On top of the virtual network where the UAU process takes place, nodes spread the awareness of the epidemics. The states in this process are unaware (U), and aware (A) of the existence of the epidemics and its prevention. Unaware individuals do not have information about how to prevent infection, while
aware individuals reduce their risk to be infected. Awareness can come from two sources, the communication with aware neighbors (becoming aware with a probability $\lambda$) or because the individual is already infected. Since the awareness corresponds to cycles parallel to the seasonality of the epidemics, there is a certain probability of an individual to forget the awareness or not to care about it, and become again, at all effects, unaware (with a probability $\delta$).

In the physical layer, the nodes are susceptible (S) or infected (I). The infection propagates from certain infected individuals to their neighbors with a probability $\beta$, and infected nodes eventually recover with probability $\mu$. After an individual gets infected it is automatically aware of the infection and changes its state in the virtual contact layer. On the other hand, if an individual is aware in the virtual layer and is  susceptible in the physical layer, it reduces its own infectivity by a factor $\gamma$. We distinguish between the original unaware infectivity $\beta^{\mbox \scriptsize U}$ and the subsequent infectivity after being aware of the infection $\beta^{A}=\gamma\beta^{\mbox \scriptsize U}$. In the particular case of $\gamma=0$, the aware individuals are completely immune to the infection.

According to this scheme, an individual can be in three different states: unaware and susceptible (US), aware and susceptible (AS) or aware and infected (AI). Note that the state unaware and infected (UI) is spurious because according the definition of the dynamical process stated it becomes immediately (AI). We propose the use of probability trees to reveal the possible states of the nodes and their transitions, see scheme in Fig.~\ref{fig2}. The MMCA equations for the coupled dynamics in the multiplex are derived using the total probability of the different states according to Fig.~\ref{fig2}.

Let us denote $a_{ij}$ and $b_{ij}$ the adjacency matrices that support the UAU and the SIS processes, respectively. Every node $i$ has a certain probability of being in one of the three states at time $t$, denoted by $p_i^{\mbox{\scriptsize{AI}}}(t)$, $p_i^{\mbox{\scriptsize{AS}}}(t)$, and $p_i^{\mbox{\scriptsize{US}}}(t)$ respectively. Assuming the absence of dynamical correlations \cite{boguna09}, the transition probabilities for node $i$ not being informed by any neighbors $r_i(t)$, not being infected by any neighbors if $i$ was aware $q_{i}^{\mbox{\scriptsize{A}}}(t)$, and not being infected by any neighbors if $i$ was unaware $q_{i}^{\mbox{\scriptsize{U}}}(t)$ are

\begin{eqnarray}
\nonumber
r_{i} (t)&=& \prod_j (1-a_{ji}p_j^{\mbox{\scriptsize{A}}}(t) \lambda)\\
\nonumber
q_{i}^{\mbox{\scriptsize{A}}}(t)&=& \prod_j (1-b_{ji}p_j^{\mbox{\scriptsize{AI}}}(t) \beta^{\mbox{\scriptsize{A}}})\\
%\nonumber
q_{i}^{\mbox{\scriptsize{U}}}(t)&=& \prod_j (1-b_{ji}p_j^{\mbox{\scriptsize{AI}}}(t) \beta^{\mbox{\scriptsize{U}}})
\label{cont}
\end{eqnarray}
\noindent where $p_j^{\mbox{\scriptsize{A}}} = p_j^{\mbox{\scriptsize{AI}}}+p_j^{\mbox{\scriptsize{AS}}}$.
Using Eqs.~\ref{cont} and the scheme presented in Fig.~\ref{fig2} we can develop the Microscopic Markov Chains for the coupled processes for each node $i$ as
\begin{widetext}
\begin{eqnarray}
%\nonumber
p_i^{\mbox{\scriptsize{US}}}(t+1)&=& p_i^{\mbox{\scriptsize{AI}}}(t)\delta\mu + p_i^{\mbox{\scriptsize{US}}}(t) r_i(t)q_i^{\mbox{\scriptsize{U}}}(t) + p_i^{\mbox{\scriptsize{AS}}}\delta q_i^{\mbox{\scriptsize{U}}}(t)
\label{pis}\\
\nonumber
p_i^{\mbox{\scriptsize{AS}}}(t+1)&=& p_i^{\mbox{\scriptsize{AI}}}(t)(1-\delta)\mu
%\nonumber
+ p_i^{\mbox{\scriptsize{US}}}(1- r_i(t))q_i^{\mbox{\scriptsize{A}}}(t) + p_i^{\mbox{\scriptsize{AS}}}(t)(1-\delta)q_i^{\mbox{\scriptsize{A}}}(t) \\
\nonumber
p_i^{\mbox{\scriptsize{AI}}}(t+1) &=& p_i^{\mbox{\scriptsize{AI}}}(t)(1-\mu)  + p_i^{\mbox{\scriptsize{US}}}\left[(1- r_i(t))(1-q_i^{\mbox{\scriptsize{A}}}(t)) + r_i(t)(1-q_i^{\mbox{\scriptsize{U}}}(t)) \right]
%\nonumber
+p_i^{\mbox{\scriptsize{AS}}}(t)\left[\delta(1-q_i^{\mbox{\scriptsize{U}}}(t))+ (1-\delta)(1-q_i^{\mbox{\scriptsize{A}}}(t)) \right]
\end{eqnarray}
\end{widetext}

The stationary solution of the system of Eqs.~\ref{pis} is computed as a set of fixed point equations satisfying $p_i^{\mbox{\scriptsize{AI}}}(t+1)=p_i^{\mbox{\scriptsize{AI}}}(t)=p_i^{\mbox{\scriptsize{AI}}}$ and equivalently for (US) and (AS). Using stationarity we are now in the position of computing the onset of the epidemics $\beta_c$. Near the critical point the MMCA can be expanded assuming that the probability of nodes to be infected in the physical layer is $p_i^{\mbox{\scriptsize{AI}}}=\epsilon_i\ll1$. Consequently, $q_{i}^{\mbox{\scriptsize{A}}}\approx 1 - \beta^{\mbox{\scriptsize{A}}} \sum_j  b_{ji}\epsilon_j$ and  $q_{i}^{\mbox{\scriptsize{U}}}\approx 1 - \beta^{\mbox{\scriptsize{U}}} \sum_j  b_{ji}\epsilon_j$. Inserting this in Eqs.~\ref{pis} we obtain
\begin{eqnarray}
%\nonumber
p_i^{\mbox{\scriptsize{US}}}&=& p_i^{\mbox{\scriptsize{US}}} r_i + p_i^{\mbox{\scriptsize{AS}}}\delta
\label{pisaprox}\\
\nonumber
p_i^{\mbox{\scriptsize{AS}}}&=& p_i^{\mbox{\scriptsize{US}}}(1- r_i)+ p_i^{\mbox{\scriptsize{AS}}}(1-\delta) \\
\nonumber
\mu \epsilon_i &=& \left( p_i^{\mbox{\scriptsize{AS}}} \beta^{\mbox{\scriptsize{A}}}  + p_i^{\mbox{\scriptsize{US}}} \beta^{\mbox{\scriptsize{U}}}\right) \sum_j  b_{ji}\epsilon_j
\end{eqnarray}
\noindent and therefore
\beq
\sum_{j} \left[
\left(1 - (1-\gamma) p_i^{\mbox{\scriptsize{A}}} \right)b_{ji} -\frac{\mu}{\beta^{\mbox{\scriptsize{U}}}}\delta_{ji}
\right]\epsilon_j =0
\label{suma}
\eeq
\noindent where $\delta_{ij}$ are the elements of the identity matrix. Note that the solution of Eq.~\eqref{suma} reduces to an eigenvalue problem for the matrix $H$ whose elements are $h_{ji} = (1 - (1-\gamma) p_i^{\mbox{\scriptsize{A}}})b_{ji} $. The onset of the epidemics is the minimum value of $\beta^{\mbox{\scriptsize{U}}}$ satisfying Eq.~\eqref{suma}. Denoting $\Lambda_{\mbox{\scriptsize{max}}}(H)$ the largest eigenvalue of $H$, the critical point is written as
\beq
\beta_c^{\mbox{\scriptsize{U}}}=\frac{\mu}{\Lambda_{\mbox{\scriptsize{max}}}(H)}
\label{critic}
\eeq

Note that $\beta_c$ depends explicitly on the dynamics on the virtual layer, in particular of the values of $p_i^{\mbox{\scriptsize{A}}}$. Interestingly, if  we consider the critical value $\lambda_c=\delta/\Lambda_{\mbox{\scriptsize{max}}}(A)$ of the onset of awareness when decoupled from the infection, i.e.\ as a simple spreading process on the virtual layer with no interaction with the physical layer, then for $\lambda<\lambda_c$ Eq.~\eqref{critic} reduces to $\beta_c=\mu/\Lambda_{\mbox{\scriptsize{max}}}(B)$, and the onset of the epidemics, is obviously independent of the awareness. The point $(\lambda_c,\beta_c)$ defines a sort of {\em meta-critical} point for the epidemic spreading.
\rerev{It is worth mentioning that this point could be a tricritical point because even though there are only two different phases in the steady state, those corresponding to the classical SIS, in the transient, for certain values of beta, there is an initial amplification of the number of infectious nodes. Later on, the awareness level increases and the infection level goes back down towards extinction.}
For values of $\lambda>\lambda_c$ the onset of the epidemics depends on the structure of the virtual layer and the dynamics of the awareness.  Specifically, it depends on the stationary values of the probabilities $p_i^{\mbox{\scriptsize{A}}}$ of the virtual layer, decoupled from the multiplex. These values are found by solving the fixed point equations of the virtual layer only.
\begin{figure}[tbp]
\begin{center}
  \includegraphics[width=\columnwidth]{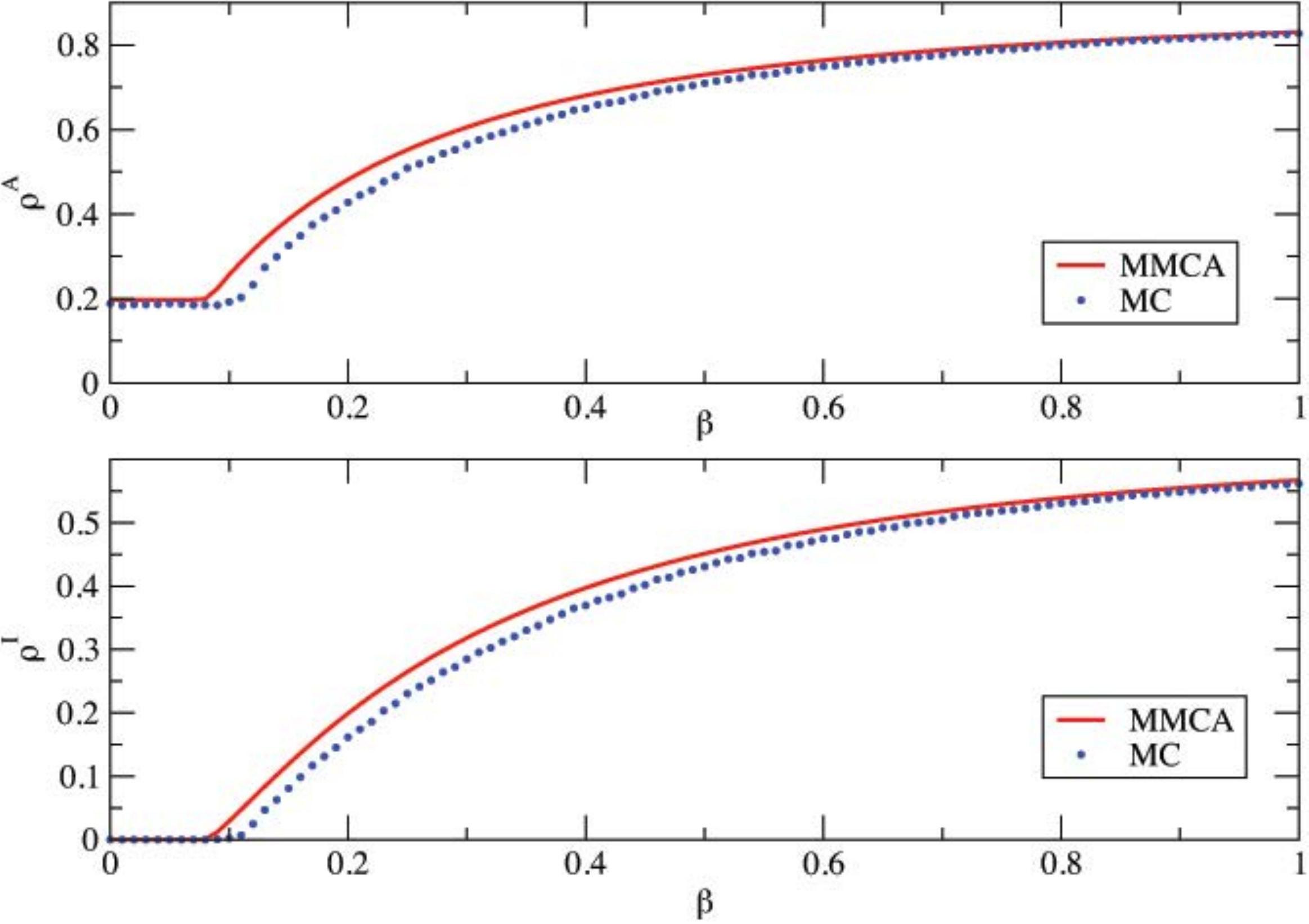}
\caption{(color online) Top: comparison of the stationary fraction of aware individuals $\rho^A=\frac{1}{N}\sum_i p_i^{A}$ using Monte Carlo (dotted line) simulations and the MMCA approach (solid line) as a function of the infectivity $\beta$ for a fixed value of $\lambda = 0.15$. Bottom: comparison of the stationary fraction of infected individuals $\rho^I=\frac{1}{N}\sum_i p_i^{I}$ using Monte Carlo (dotted line) simulations and the MMCA approach (solid line) as a function of the infectivity $\beta$.
The initial fraction of infected nodes is set to 0.2. The multiplex structure is, in this case: i) physical layer, a scale-free network of 1000 nodes generated with the configurational model, and with exponent $2.5$, ii) virtual layer, the same network than in the physical layer but with 400 extra random links (non-overlapping with previous). The values for the recovery probabilities are $\delta=0.6$, and $\mu=0.4$.}
\label{fig3}
\end{center}
\end{figure}

We crosscheck our analytical results with extensive computer simulations of the coupled dynamics UAU-SIS in different configurations of multiplex. For the sake of simplicity, we will present the results for $\gamma=0$, meaning that $\beta^{\mbox{\scriptsize{A}}}=0$ (and henceforth $q_i^{A}=1$, and $\beta^{\mbox{\scriptsize{U}}} =\beta$).This corresponds to complete immunity of nodes aware of the infection, although the calculation is identical for any other different value of $\gamma$.

In Fig.~\ref{fig3} we plot the comparison of MMCA with Monte Carlo simulations, for a quenched multiplex of two layers, in the physical layer we build a power-law degree distribution network generated with the configurational model with exponent $2.5$ of 1000 nodes, and in the virtual layer the same network with 400 extra random links (non-overlapping with previous). Note that the MMCA approach is specially suited for quenched networks, and then it is not necessary to assess the validity of the approximation in the thermodynamic limit \cite{gomez10,gomez11}.
The average accuracy of the approximation is $\sim 2\%$. We use this multiplex as a representation of a structure that could account for a realistic scenario, where all the connections in the physical layer are also present in the virtual network, and the virtual network has additional links. Nevertheless, we have explored different multiplex for the sake of completion, and in all of them, the \rev{qualitative behavior is the same} (see Supplemental Material).

\begin{figure}[tbp]
%\begin{center}
%\begin{tabular}{cc}
  \includegraphics[width=0.95\columnwidth,clip=0]{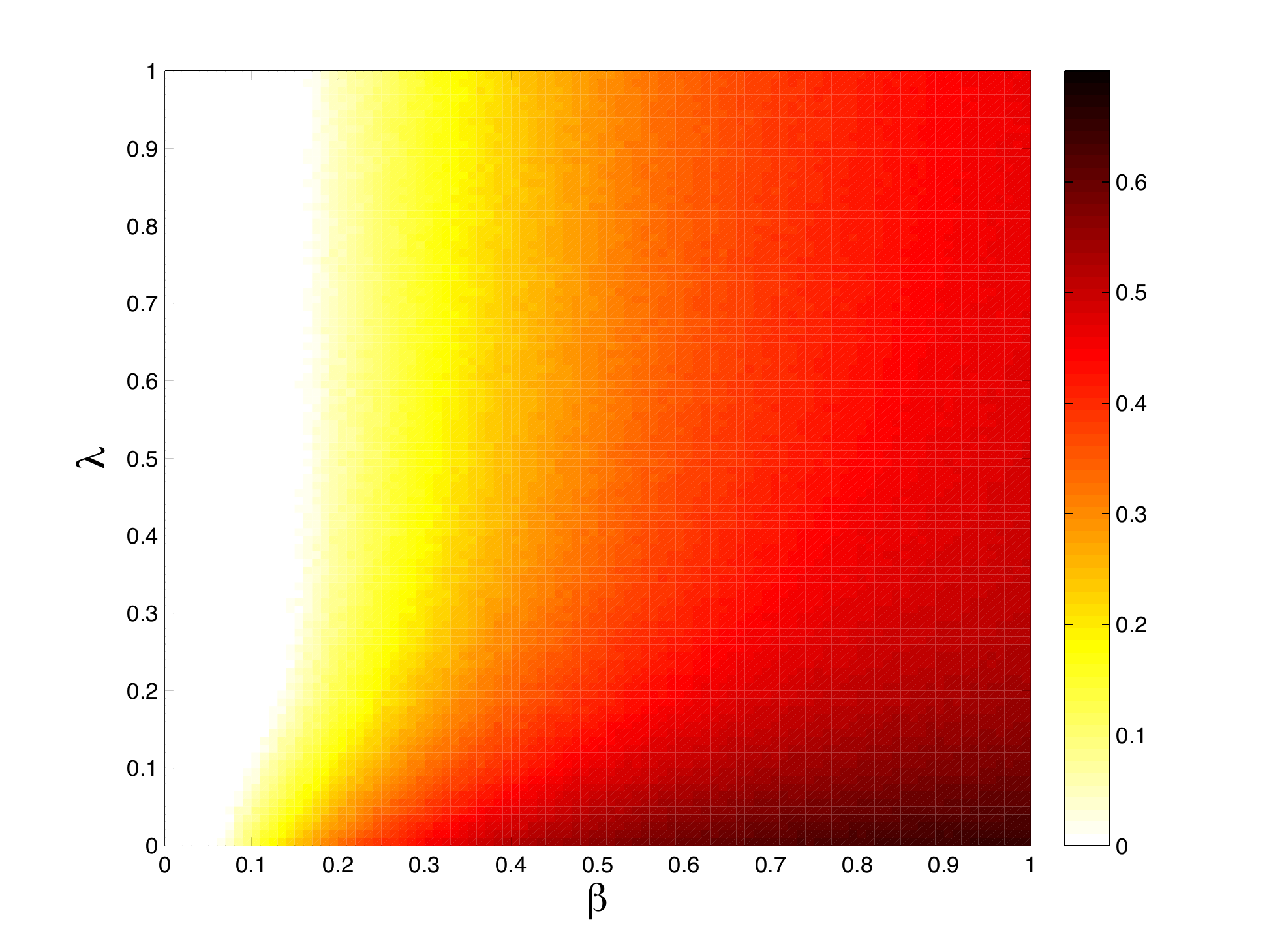}\\
  \includegraphics[width=0.95\columnwidth,clip=0]{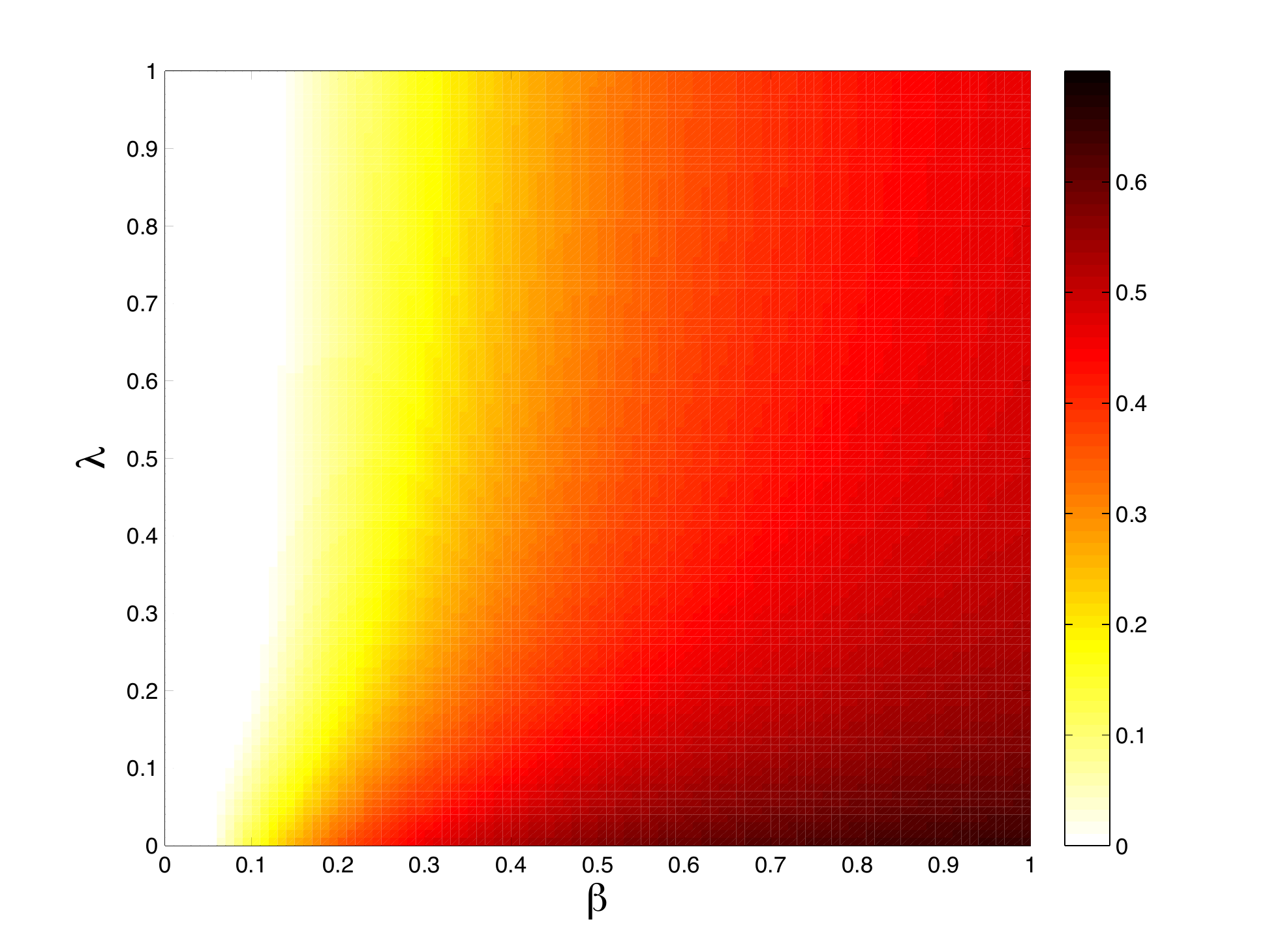}
%\end{tabular}
%\end{center}
\caption{(color online) Comparison between Monte Carlo and MMCA for the fraction $\rho^I$ of infected individuals in the stationary state (colors represent the fraction of infected individuals). Top, full phase diagram $\lambda-\beta$ for the same multiplex described in Fig.~\ref{fig3} obtained by averaging 50 Monte Carlo simulations for each point in the grid $100\times100$. Bottom, same for the MMCA. The relative error for the full phase diagram is $\approx1.6\%$.}
\label{fig4}
\end{figure}

We have also explored the full phase diagram ($\lambda-\beta$) of the dynamics UAU-SIS for the same multiplex as before, see Fig.~\ref{fig4}.  We represent the fraction of infected individuals in the whole population, in the stationary state, $\rho^I$. The agreement is very good for the full phase space, being the relative error less than $2.5\%$ in all the multiplex configurations explored, e.g.\ composing random homogeneous networks (Erd\H{o}s-R\'enyi networks) and heterogeneous networks (scale-free networks), for different values of the parameters (see Supplemental material).

\begin{figure}[tbp]
  \includegraphics[width=\columnwidth,clip=0]{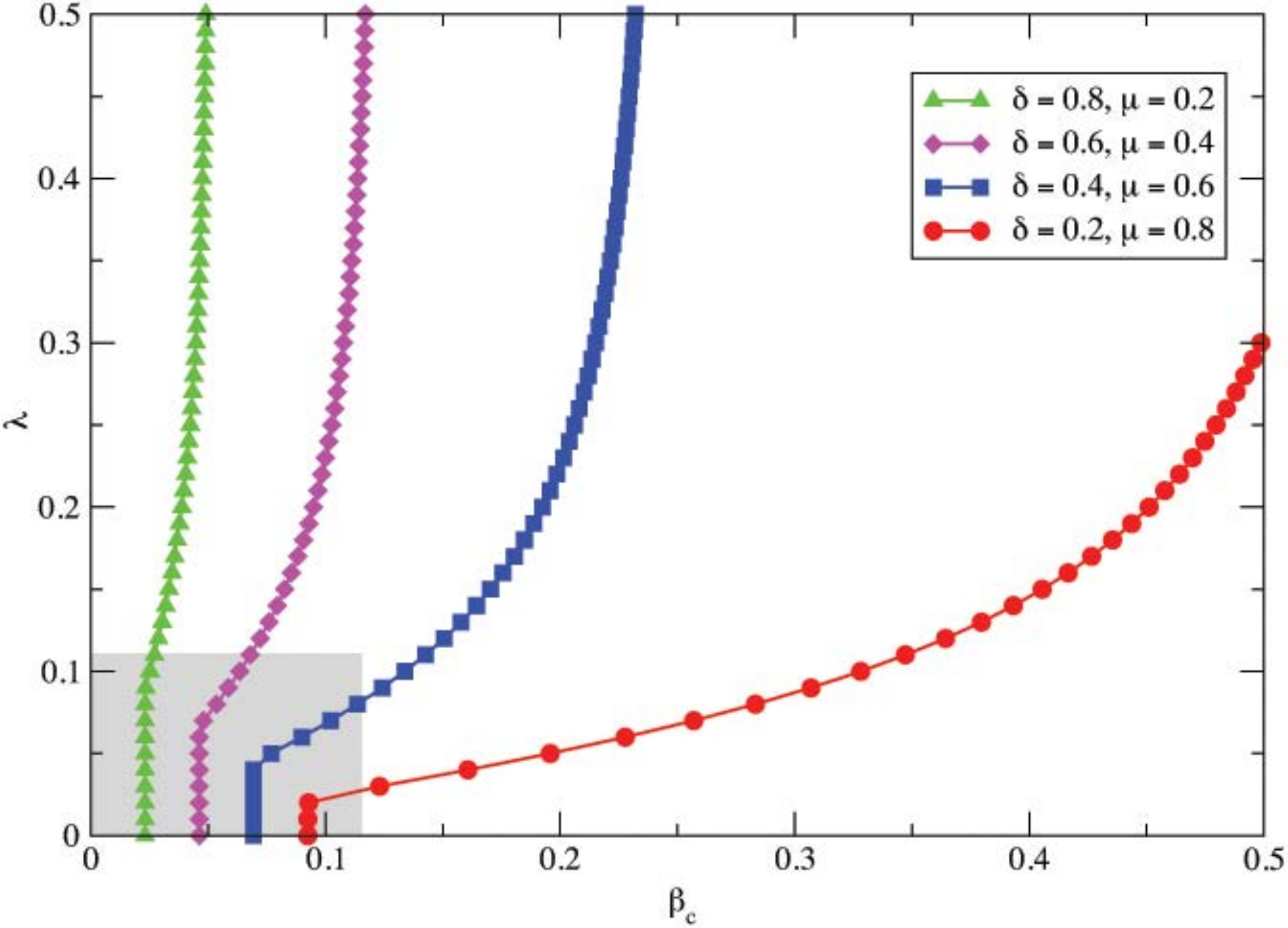}
\caption{(color online) Dependence of the onset of the epidemics $\beta_c$ as a function of $\lambda$ computed using Eq.~\eqref{critic}, for different values of the recovery $\delta$ and $\mu$, for the same multiplex described in Fig.~\ref{fig3}. The shaded rectangle corresponds to the area where the meta-critical points may be, which are bounded by the topological characteristics of the multiplex $1/\Lambda_{\mbox{\scriptsize{max}}}(A)$ and $1/\Lambda_{\mbox{\scriptsize{max}}}(B)$.}
\label{fig5}
\end{figure}

Finally, we plot the prediction of the critical epidemic threshold line $\beta_c(\lambda)$ given by Eq.~\eqref{critic} for different values of the recoveries $\delta$ and $\mu$, Fig.~\ref{fig5}. Note that there exists a region where the meta-critical point is localized, corresponding to the area bounded by $[0,1/\Lambda_{\mbox{\scriptsize{max}}}(A)] \times [0,1/\Lambda_{\mbox{\scriptsize{max}}}(B)]$. Looking at the curves in Fig.~\ref{fig5} we observe that initially the epidemic threshold does not depend on the awareness. At a certain point $\lambda_c$, what we call the meta-critical point, the epidemics is delayed and contained. This last effect will be observed for any value $(\lambda, \beta)$ outside the shaded area.
%%%%%%%%%%%%%%%%%%%%%%%%%%%%%%%%%%%%%%%%
%\section{Introduction}
%%%%%%%%%%%SUMMARY

Summarizing, we have analyzed a coupled dynamical process of awareness and infection on top of multiplex networks. The results show that the coexistence of different topologies spreading antagonistic effects raises interesting physical phenomena, as for example the emergence of a meta-critical point, where the diffusion of awareness is able to control the onset of the epidemics. Given the specific nature of the awareness spreading proposed here, equivalent to a SIS process, the results are also valid to describe two competing infectious strains coexisting in a multiplex structure, the only difference being if the strains reinforce or weaken each other. \rev{The genuine mechanism underlying the emergence of the dependence of the onset of the epidemics on the diffusion of the awareness is rooted to the cyclic character of both coupled processes. If one of the processes is not cyclic $\delta=0$ or $\mu=0$ this dependence disappears.}
The high accuracy of the MMCA is specially useful in this scenario of coupled dynamics in quenched networks, where heterogeneous mean-field approximations for binary states, or in general approximations for annealed networks \cite{romu01,newman02,gleeson11} could be difficult to define because of the structure of the multiplex, where the degree-class is multivalued.
The results provide clues to quantify the effect of the word of mouth, for example using Facebook, or twitter, in campaigns against seasonal diseases, and its power in prevention the epidemics, decrease its incidence, or eventually eradicate it.

\section{Acknowledgements}
We acknowledge Manlio De Domenico for interesting comments on the draft, \rerev{and also to the anonymous referee for his valuable comments during the referral process.}
This work has been partially supported by MINECO through Grant FIS2012-38266; and by the EC FET-Proactive Project PLEXMATH (grant 317614). A.A.\ also acknowledges partial financial support from the ICREA Academia and the James S.\ McDonnell Foundation.

%\bibliography{biblio}

%\newpage
\appendix
\section{Supplemental material}

%%%%%%%%%%%%%%%%%%%%%%%%%%%%%%%%%%%%%%%%%%%%%%%%%%%%%%%%%%%%%%%%%%%%%%%%%
\begin{figure*}[h]
  \mbox{\includegraphics*[width=0.60\textwidth]{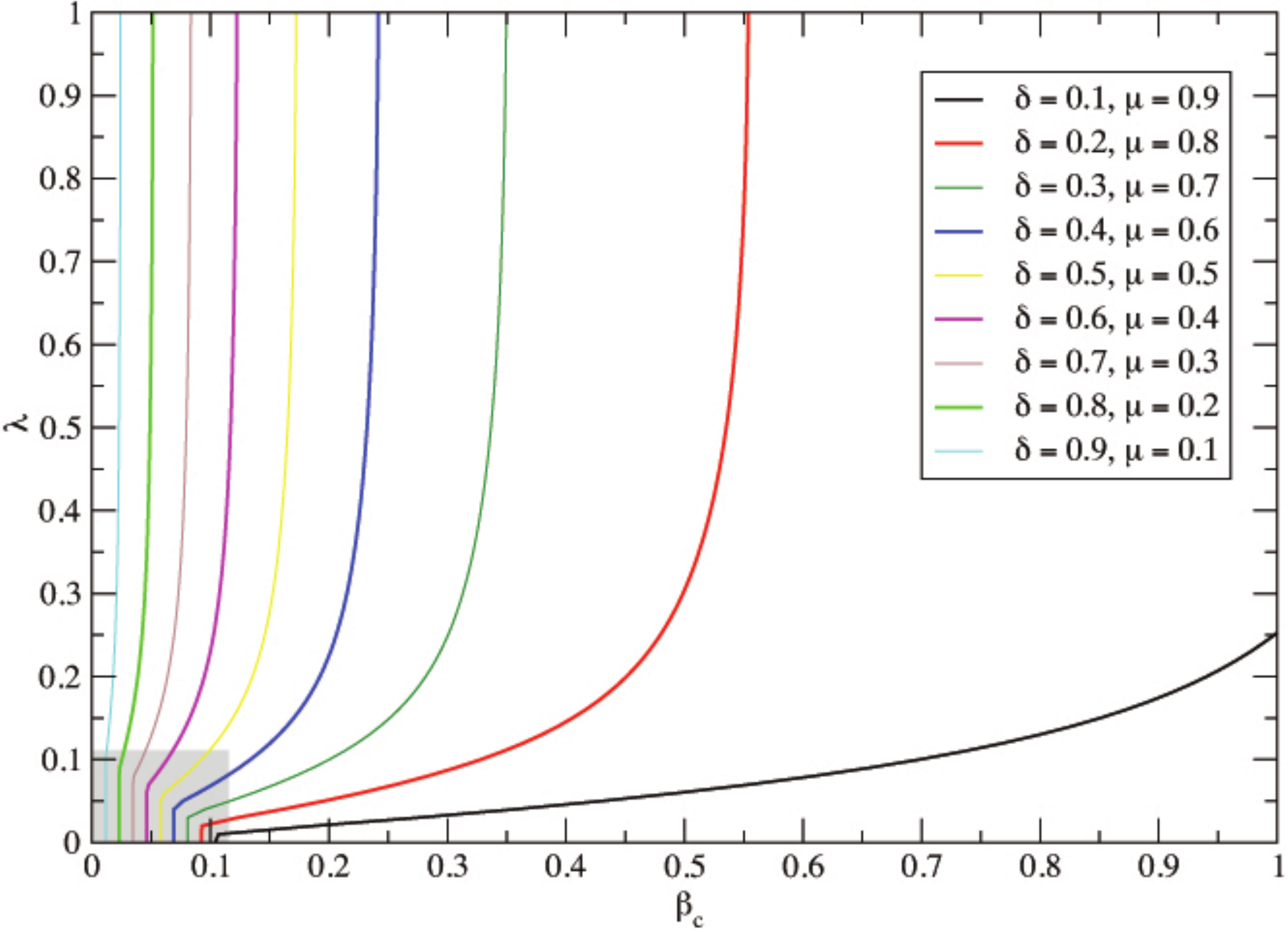}}
\caption{(color online) Dependence of the onset of the epidemics $\beta_c$ as a function of $\lambda$ computed using Eq.(5) of main text, for different values of the recovery $\delta$ and $\mu$, for a multiplex formed by: physical layer, a scale-free network of 1000 nodes with degree distribution $P(k)\sim k^{-2.5}$, and virtual layer, same scale-free network with 400 extra (non-overlapping) random links. The shaded rectangle corresponds to the area where the meta-critical points may be, which are bounded by the topological characteristics of the multiplex $1/\Lambda_{\mbox{\scriptsize{max}}}(A)$ and $1/\Lambda_{\mbox{\scriptsize{max}}}(B)$.}
%\label{fig5}
\end{figure*}

%%%%%%%%%%%%%%%%%%%%%%%%%%%%%%%%%%%%%%%%%%%%%%%%%%%%%%%%%%%%%%%%%%%%%%%
\begin{figure*}[ht]
\begin{tabular}{cc}
  MC, $\delta=0.4$, $\mu=0.6$ & MMCA, $\delta=0.4$, $\mu=0.6$ \\
  \mbox{\includegraphics*[width=0.45\textwidth,clip=0]{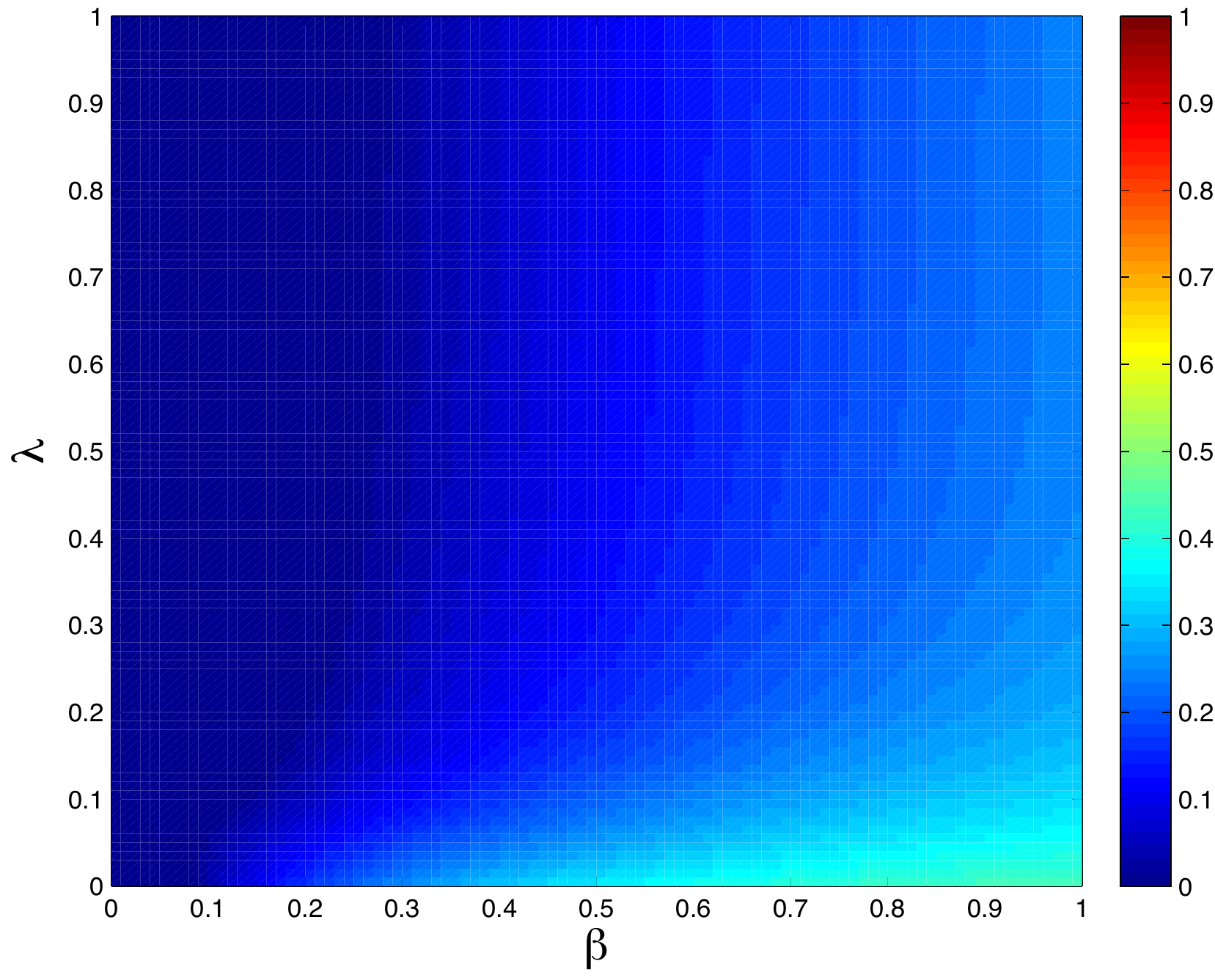}} &
  \mbox{\includegraphics*[width=0.45\textwidth,clip=0]{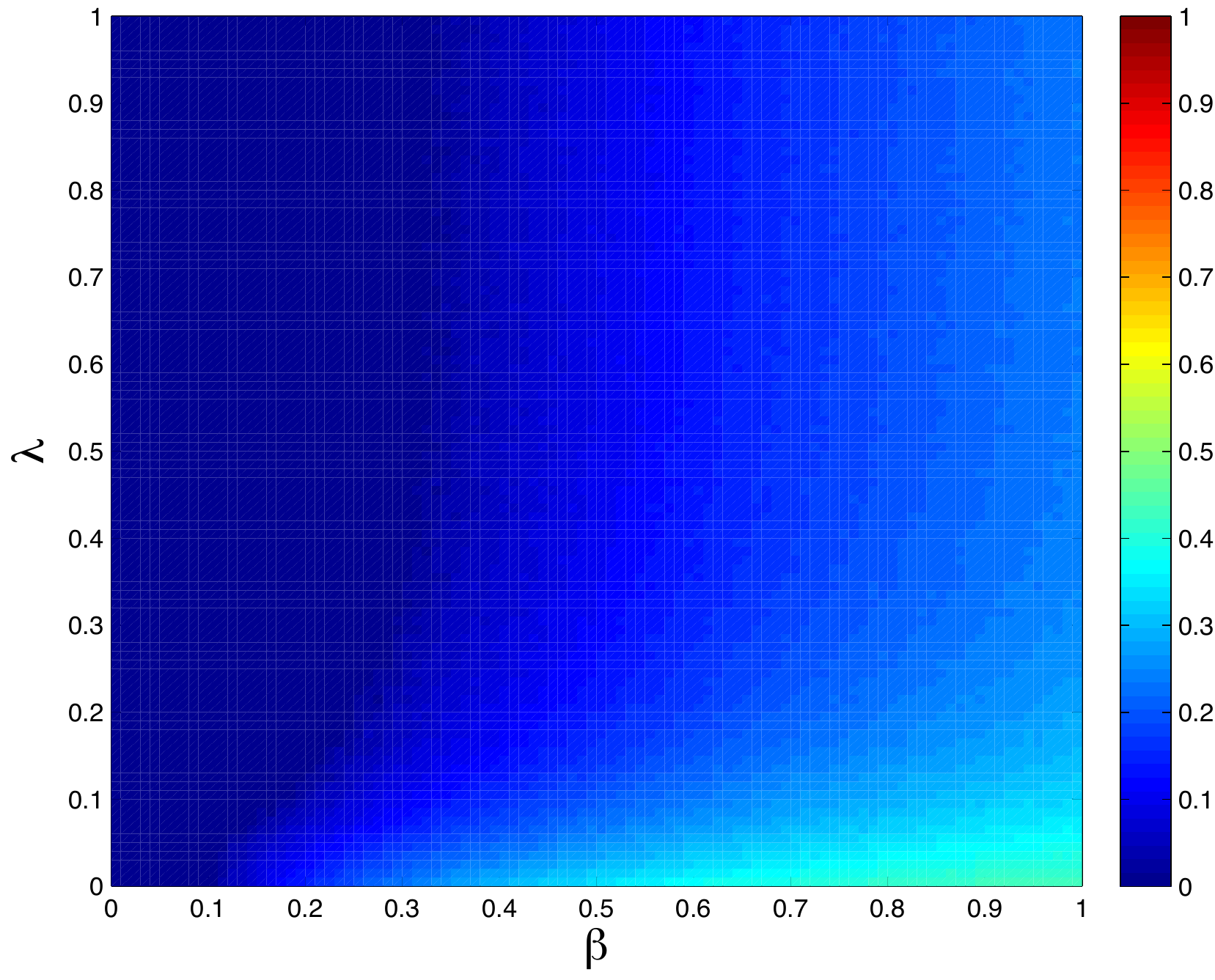}}
  \\
  MC, $\delta=0.5$, $\mu=0.5$ & MMCA, $\delta=0.5$, $\mu=0.5$ \\
  \mbox{\includegraphics*[width=0.45\textwidth,clip=0]{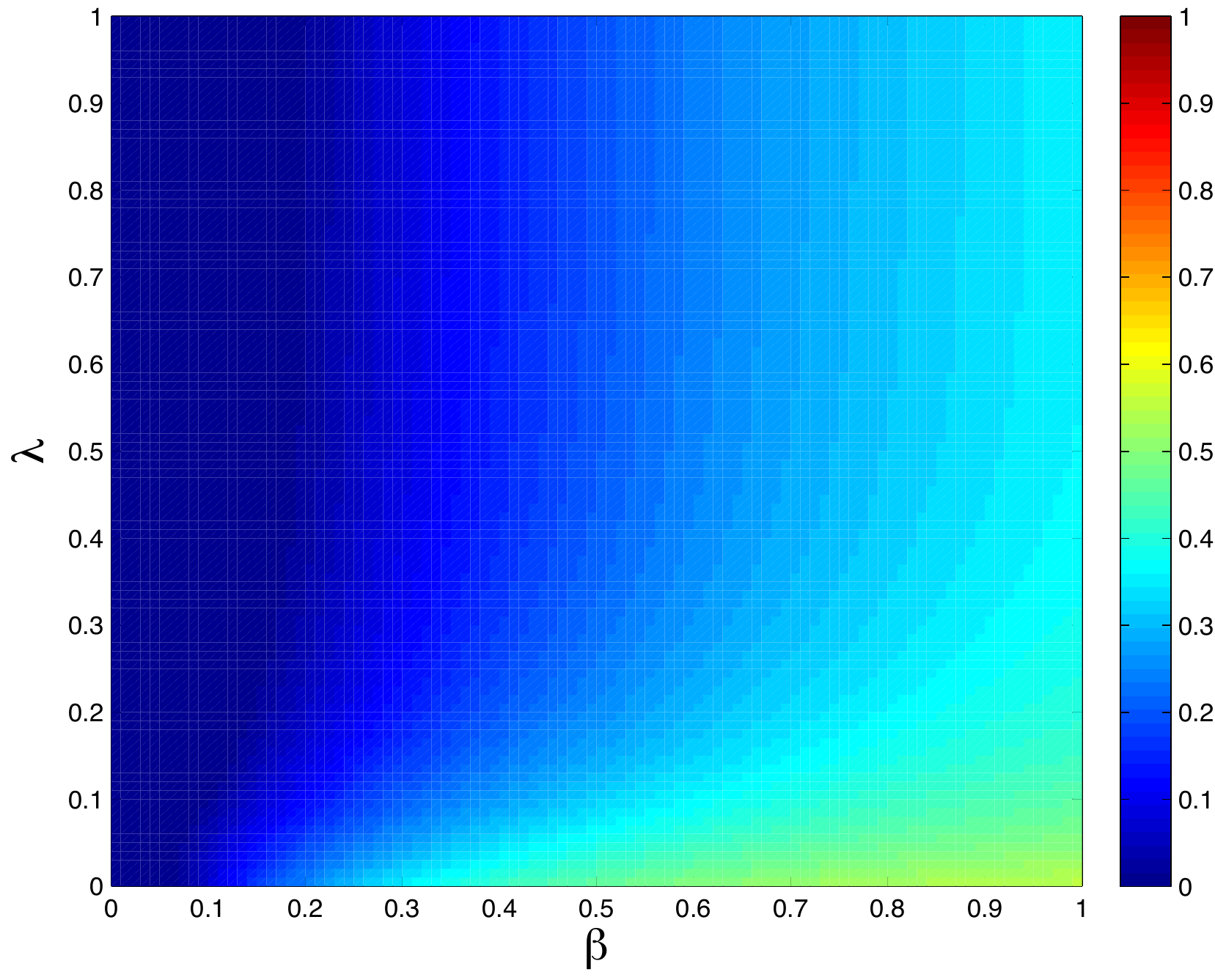}} &
  \mbox{\includegraphics*[width=0.45\textwidth,clip=0]{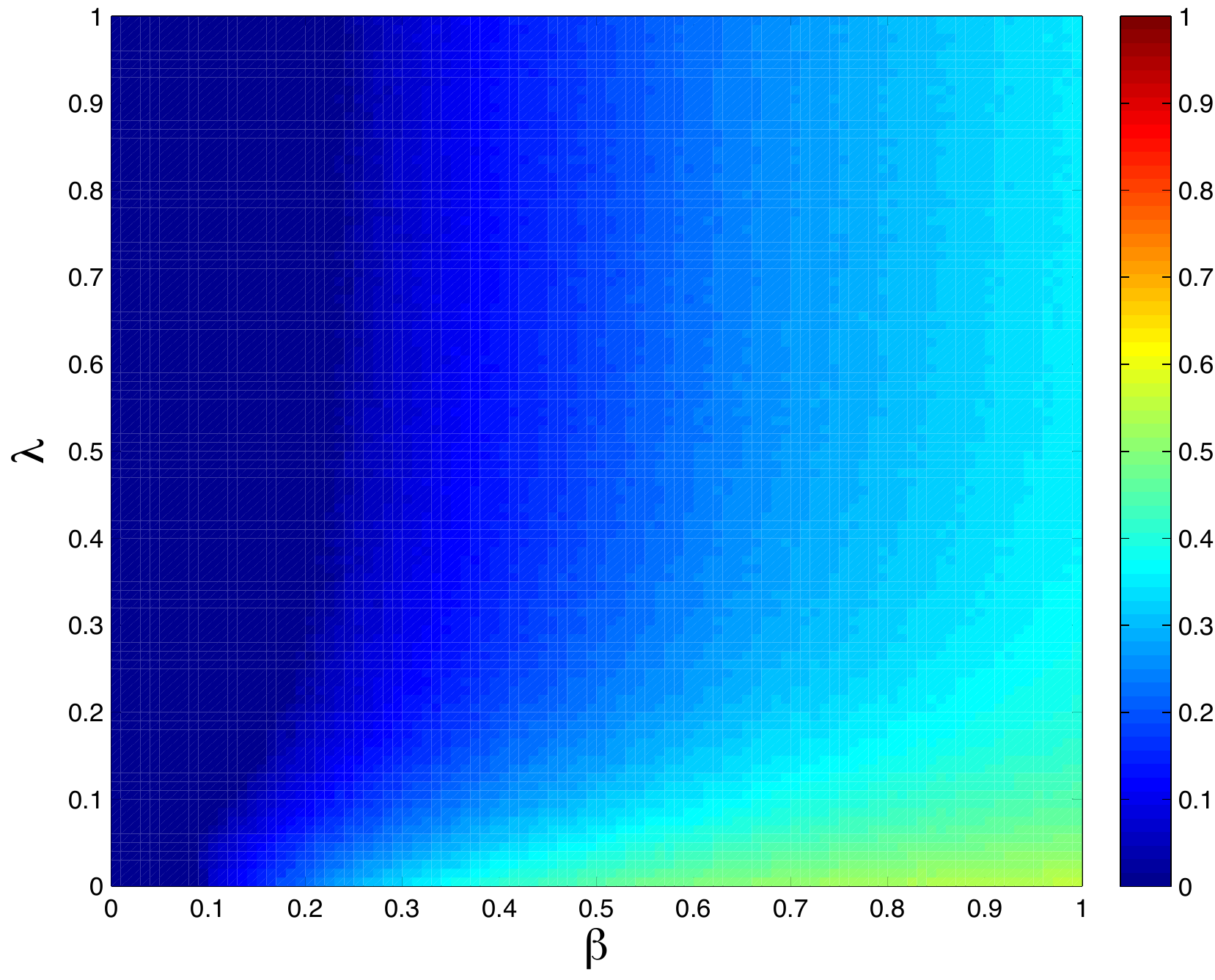}}
  \\
  MC, $\delta=0.6$, $\mu=0.4$ & MMCA, $\delta=0.6$, $\mu=0.4$ \\
  \mbox{\includegraphics*[width=0.45\textwidth,clip=0]{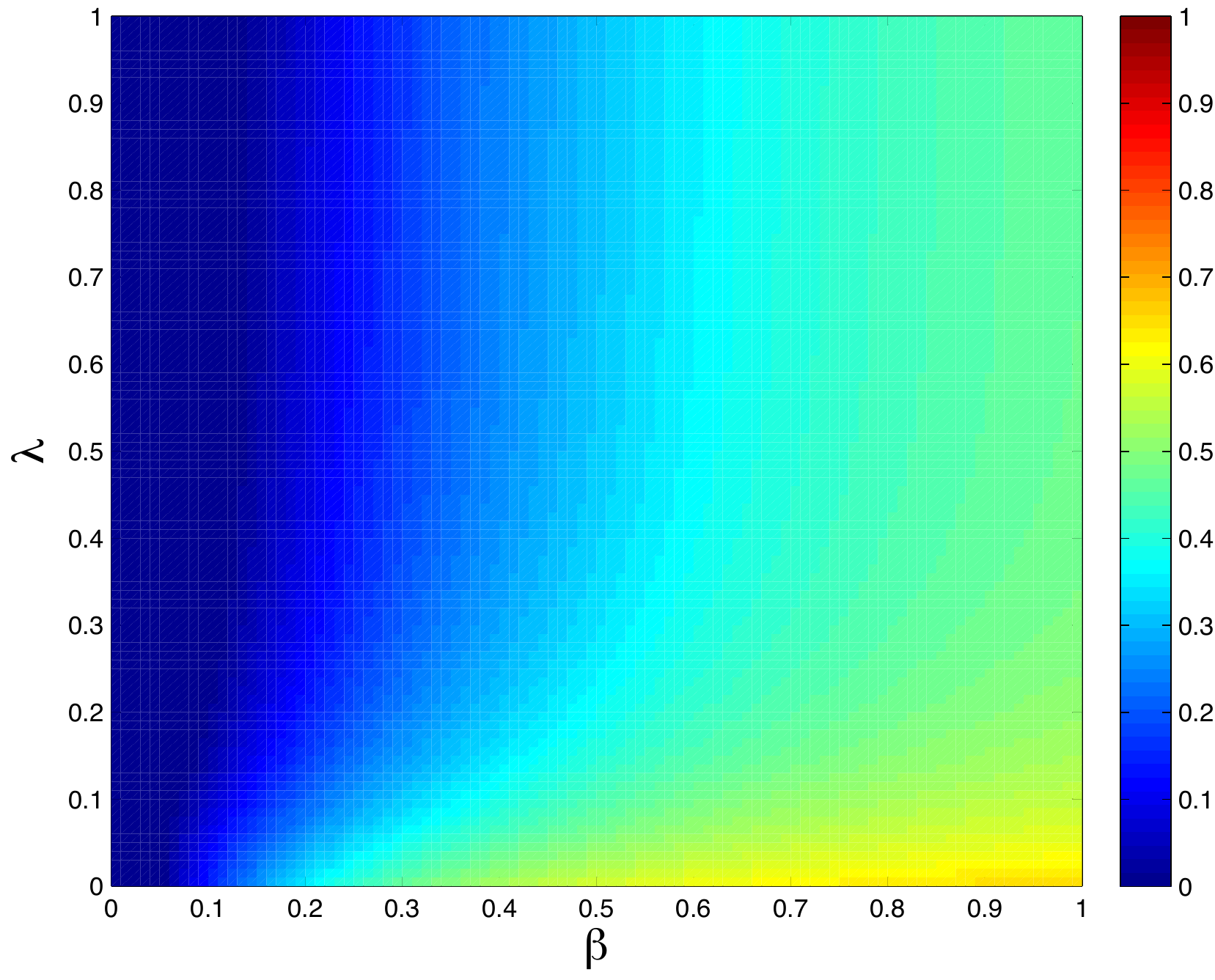}} &
  \mbox{\includegraphics*[width=0.45\textwidth,clip=0]{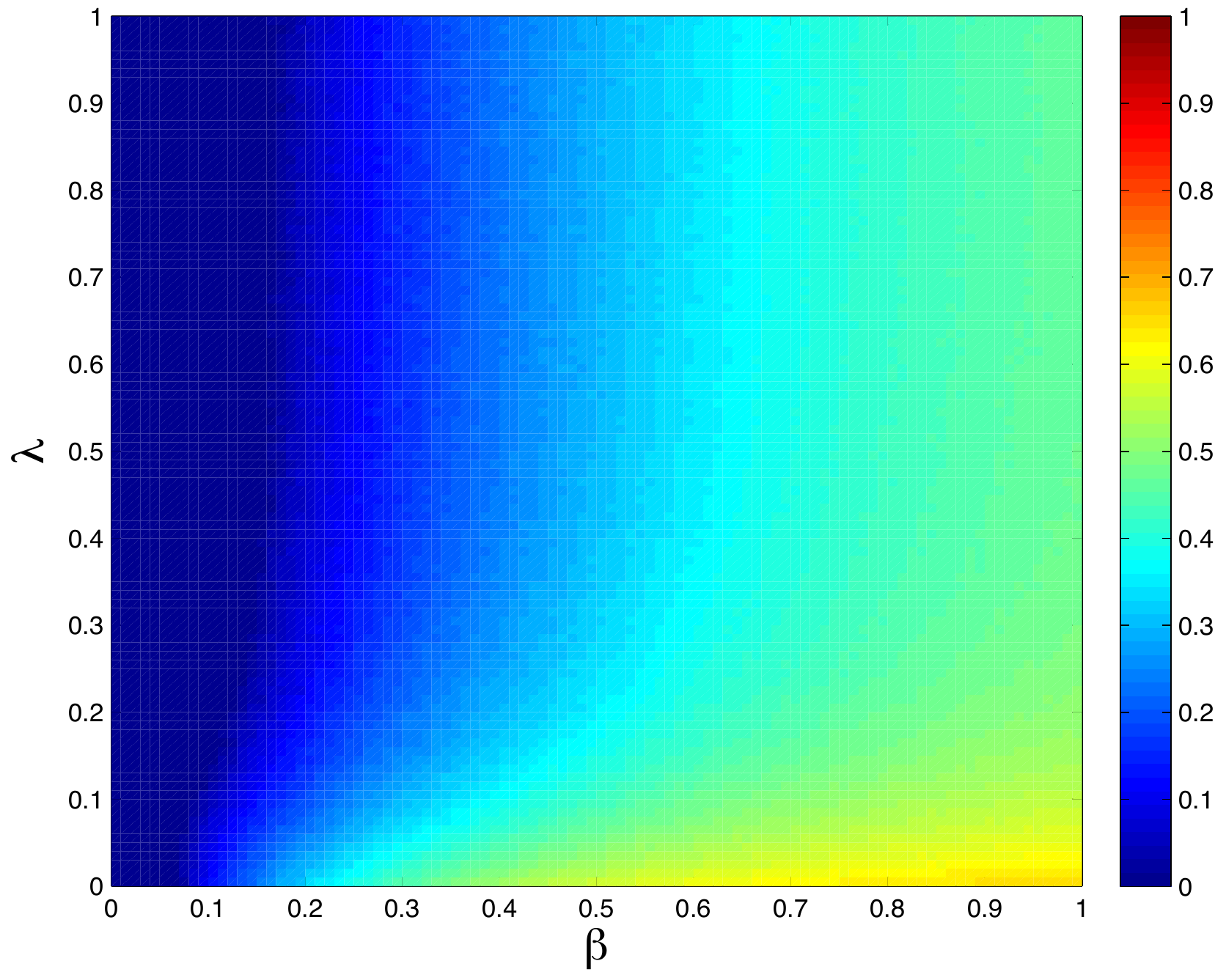}}
\end{tabular}
\caption{(color online) Comparison between Monte Carlo and MMCA for the fraction $\rho^I$ of infected individuals in the stationary state. Multiplex formed by: virtual layer, Erd\"os-R\'enyi network of 1000 nodes with $\average{k}=8$, and physical layer, a scale-free network of 1000 nodes with degree distribution $P(k)\sim k^{-2.5}$. Full $100\times100$ $\lambda-\beta$ phase diagram. MC values are averages over 50 simulations, and initial fraction of infected nodes is $20\%$. The relative errors between MC and MMCA are: $0.9\%$, $1.0\%$, and $1.2\%$, respectively.}
\end{figure*}

%%%%%%%%%%%%%%%%%%%%%%%%%%%%%%%%%%%%%%%%%%%%%%%%%%%%%%%%%%%%%%%%%%%%%%%
\begin{figure*}[ht]
\begin{tabular}{cc}
  MC, $\delta=0.4$, $\mu=0.6$ & MMCA, $\delta=0.4$, $\mu=0.6$ \\
  \mbox{\includegraphics*[width=0.45\textwidth,clip=0]{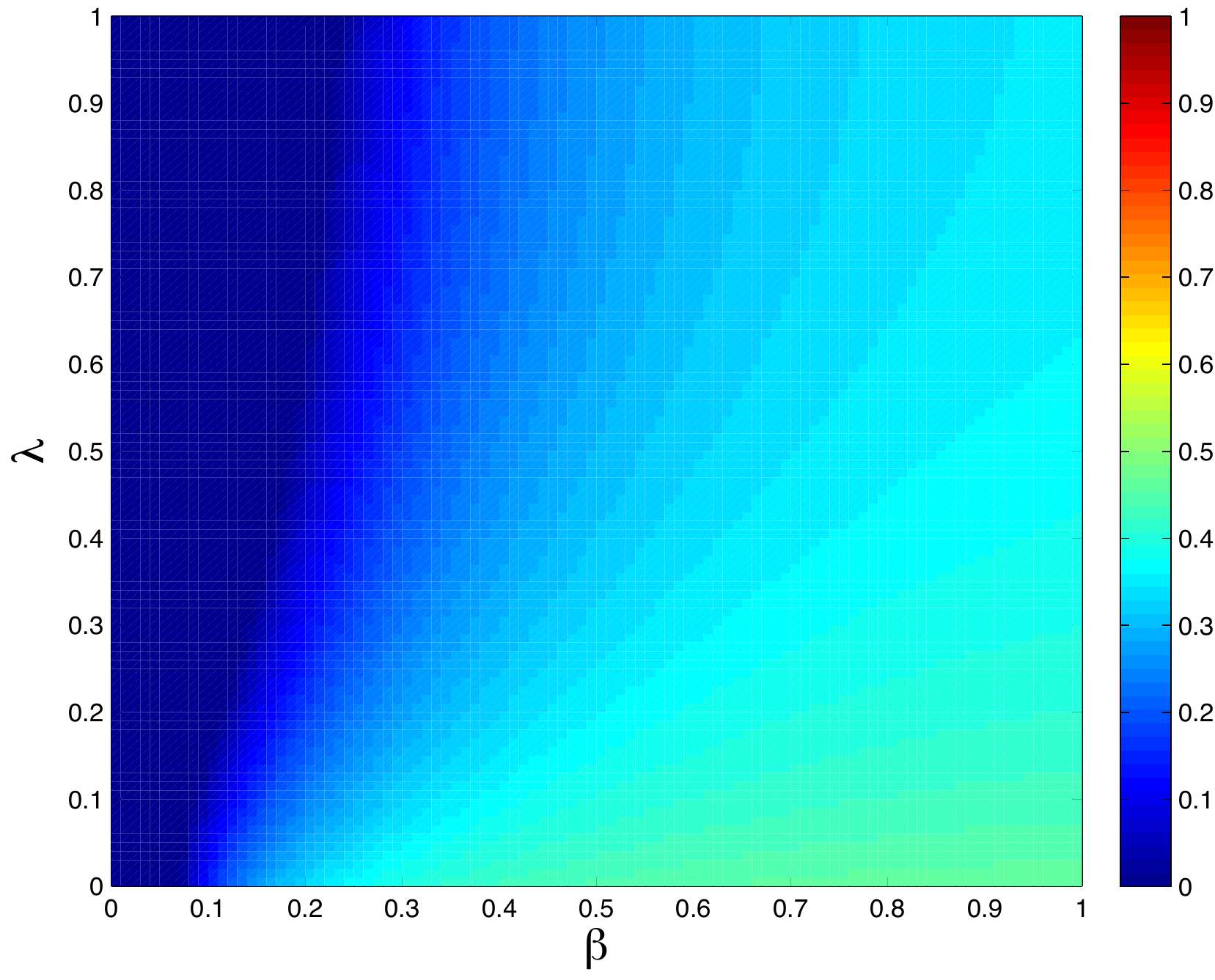}} &
  \mbox{\includegraphics*[width=0.45\textwidth,clip=0]{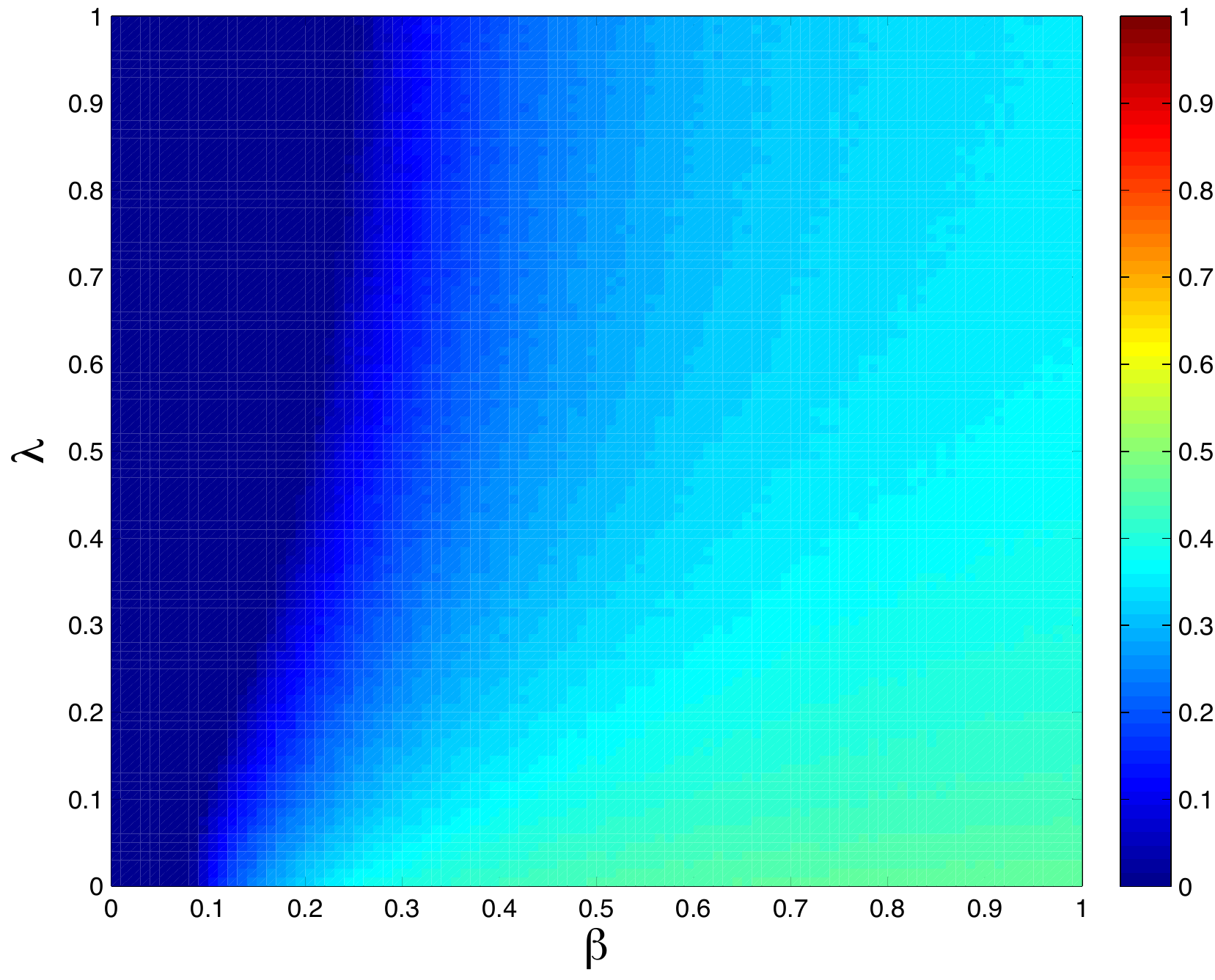}}
  \\
  MC, $\delta=0.5$, $\mu=0.5$ & MMCA, $\delta=0.5$, $\mu=0.5$ \\
  \mbox{\includegraphics*[width=0.45\textwidth,clip=0]{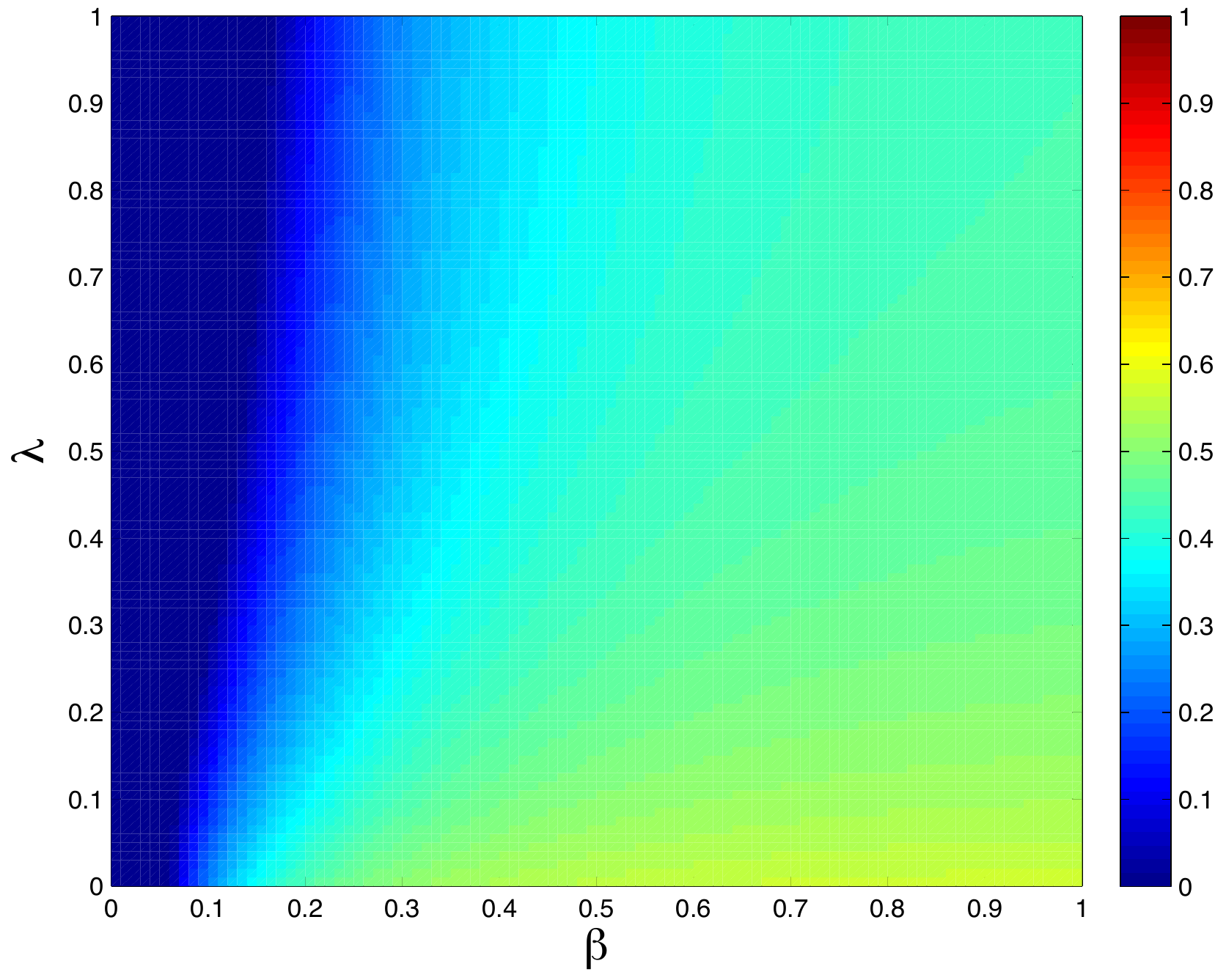}} &
  \mbox{\includegraphics*[width=0.45\textwidth,clip=0]{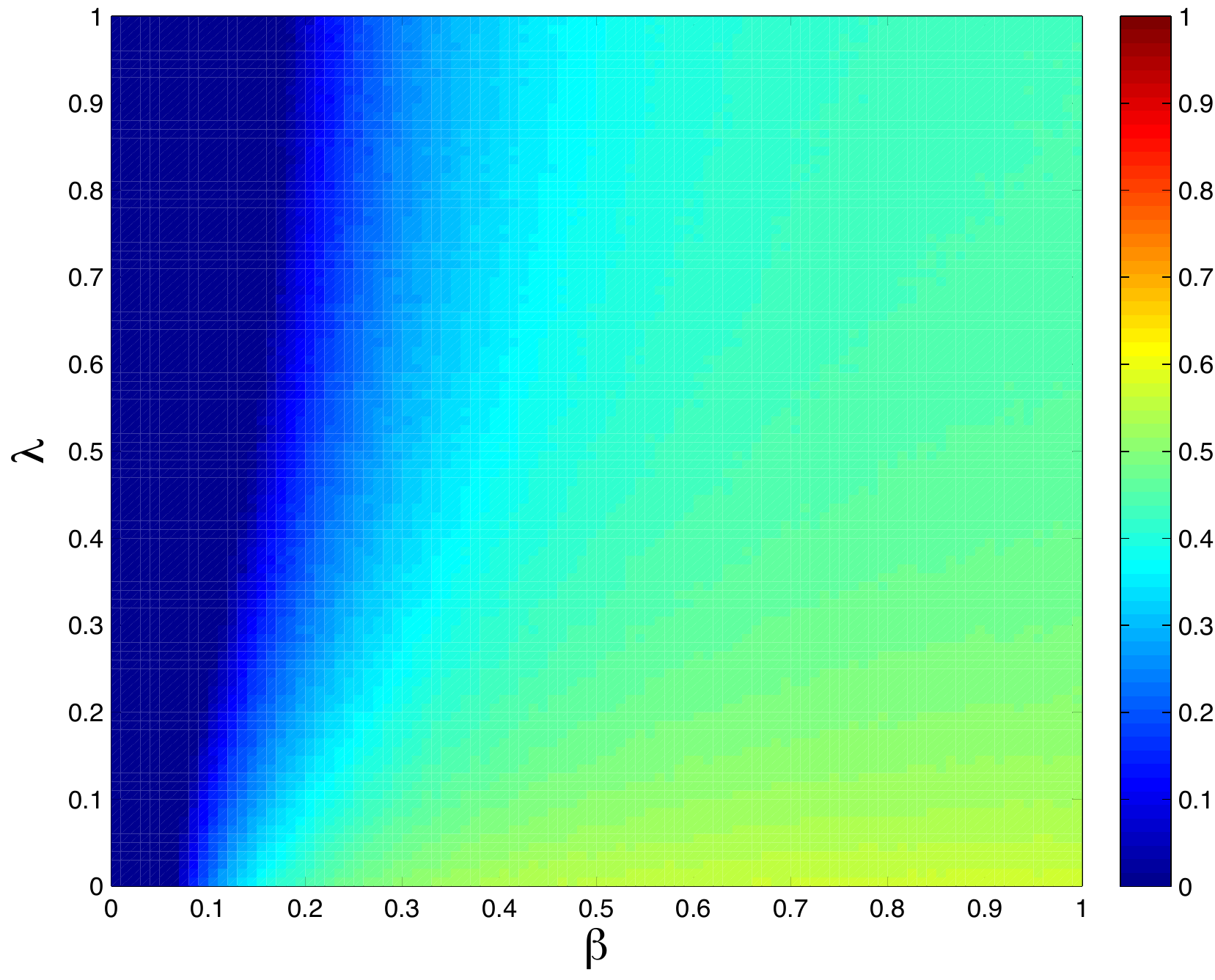}}
  \\
  MC, $\delta=0.6$, $\mu=0.4$ & MMCA, $\delta=0.6$, $\mu=0.4$ \\
  \mbox{\includegraphics*[width=0.45\textwidth,clip=0]{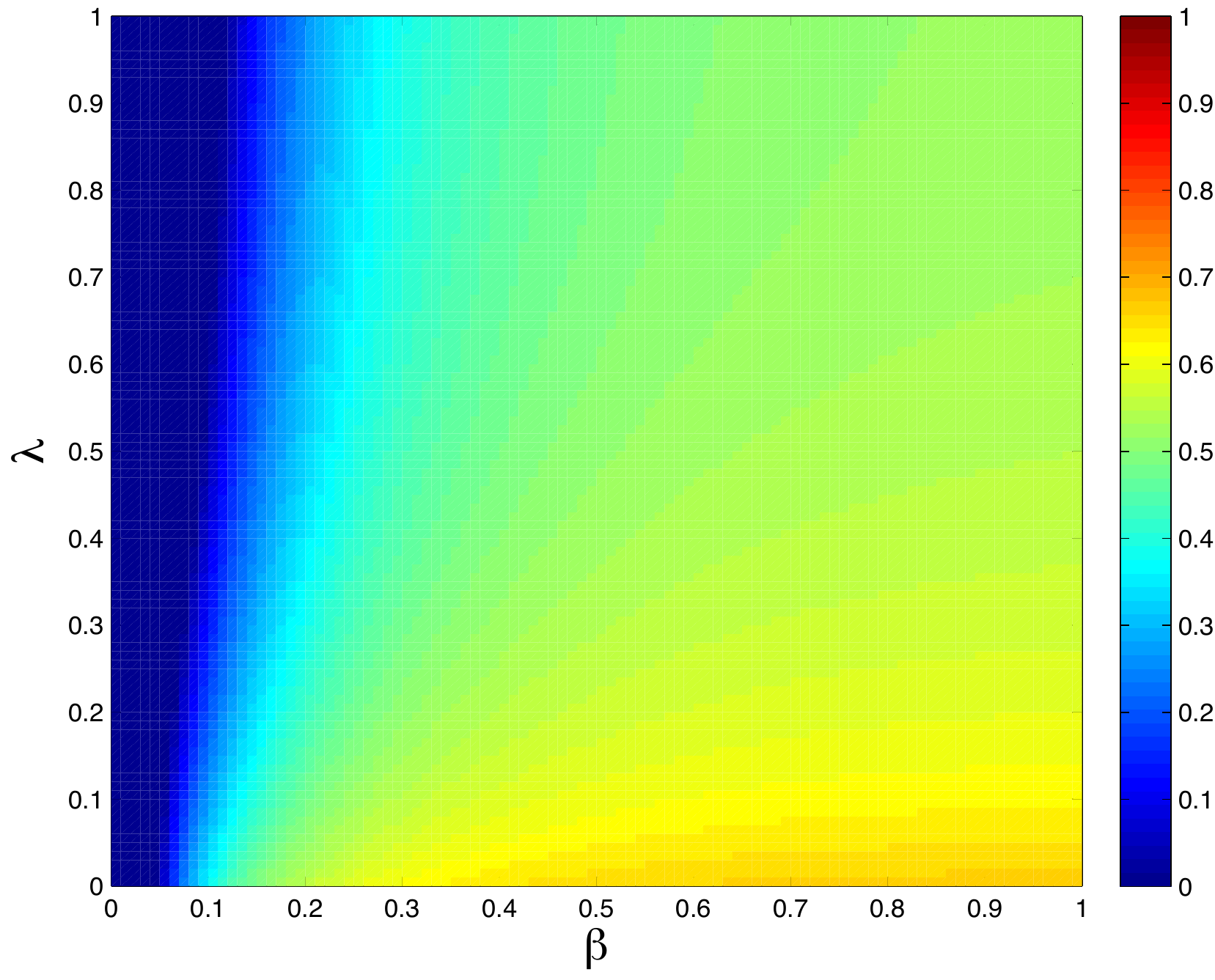}} &
  \mbox{\includegraphics*[width=0.45\textwidth,clip=0]{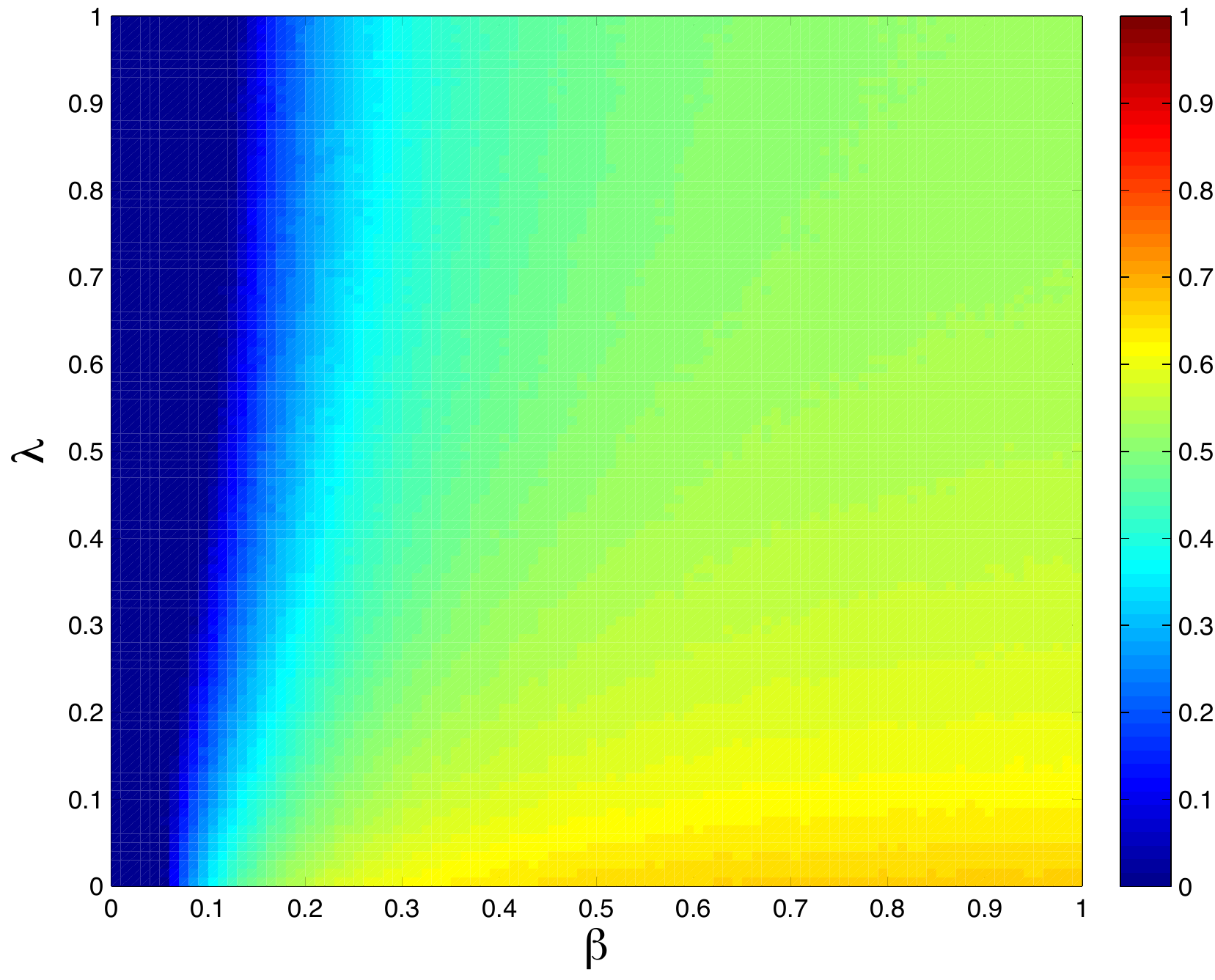}}
\end{tabular}
\caption{(color online) Comparison between Monte Carlo and MMCA for the fraction $\rho^I$ of infected individuals in the stationary state. Multiplex formed by: virtual layer, a scale-free network of 1000 nodes with degree distribution $P(k)\sim k^{-2.5}$, and physical layer, Erd\"os-R\'enyi network of 1000 nodes with $\average{k}=8$. Full $100\times100$ $\lambda-\beta$ phase diagram. MC values are averages over 50 simulations, and initial fraction of infected nodes is $20\%$. The relative errors between MC and MMCA are: $0.4\%$, $0.4\%$, and $0.4\%$, respectively.}
\end{figure*}

%%%%%%%%%%%%%%%%%%%%%%%%%%%%%%%%%%%%%%%%%%%%%%%%%%%%%%%%%%%%%%%%%%%%%%%
\begin{figure*}[ht]
\begin{tabular}{cc}
  MC, $\delta=0.4$, $\mu=0.6$ & MMCA, $\delta=0.4$, $\mu=0.6$ \\
  \mbox{\includegraphics*[width=0.45\textwidth,clip=0]{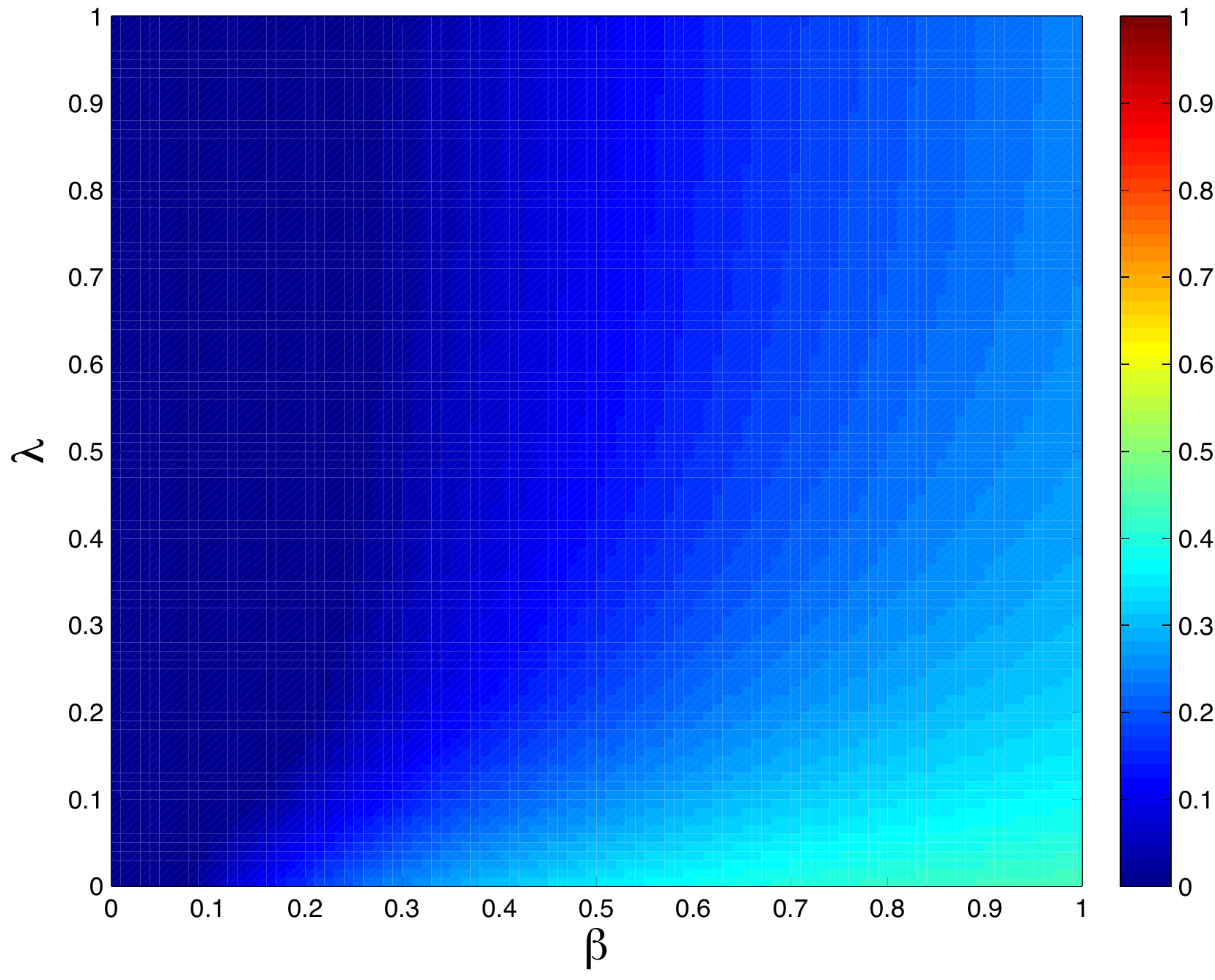}} &
  \mbox{\includegraphics*[width=0.45\textwidth,clip=0]{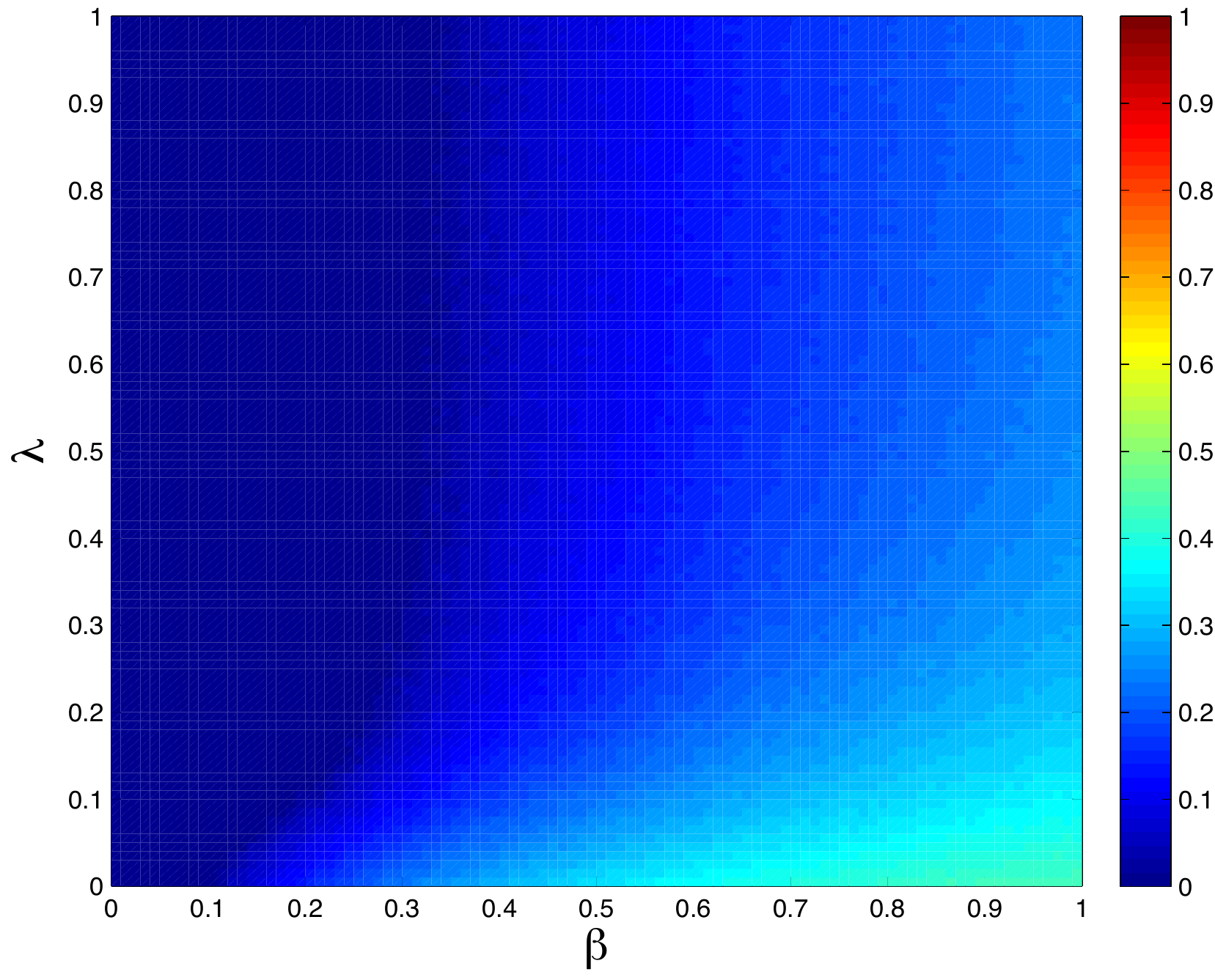}}
  \\
  MC, $\delta=0.5$, $\mu=0.5$ & MMCA, $\delta=0.5$, $\mu=0.5$ \\
  \mbox{\includegraphics*[width=0.45\textwidth,clip=0]{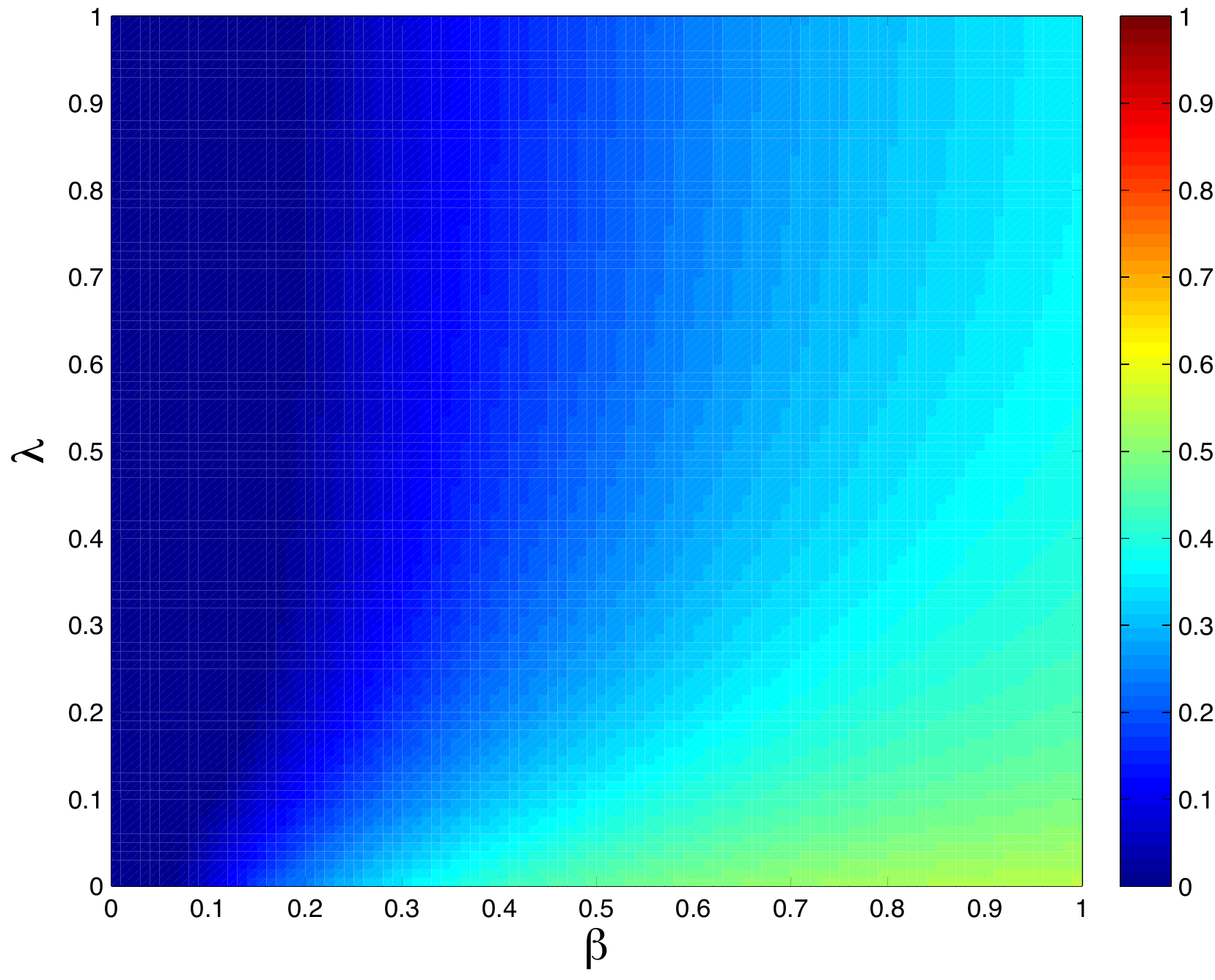}} &
  \mbox{\includegraphics*[width=0.45\textwidth,clip=0]{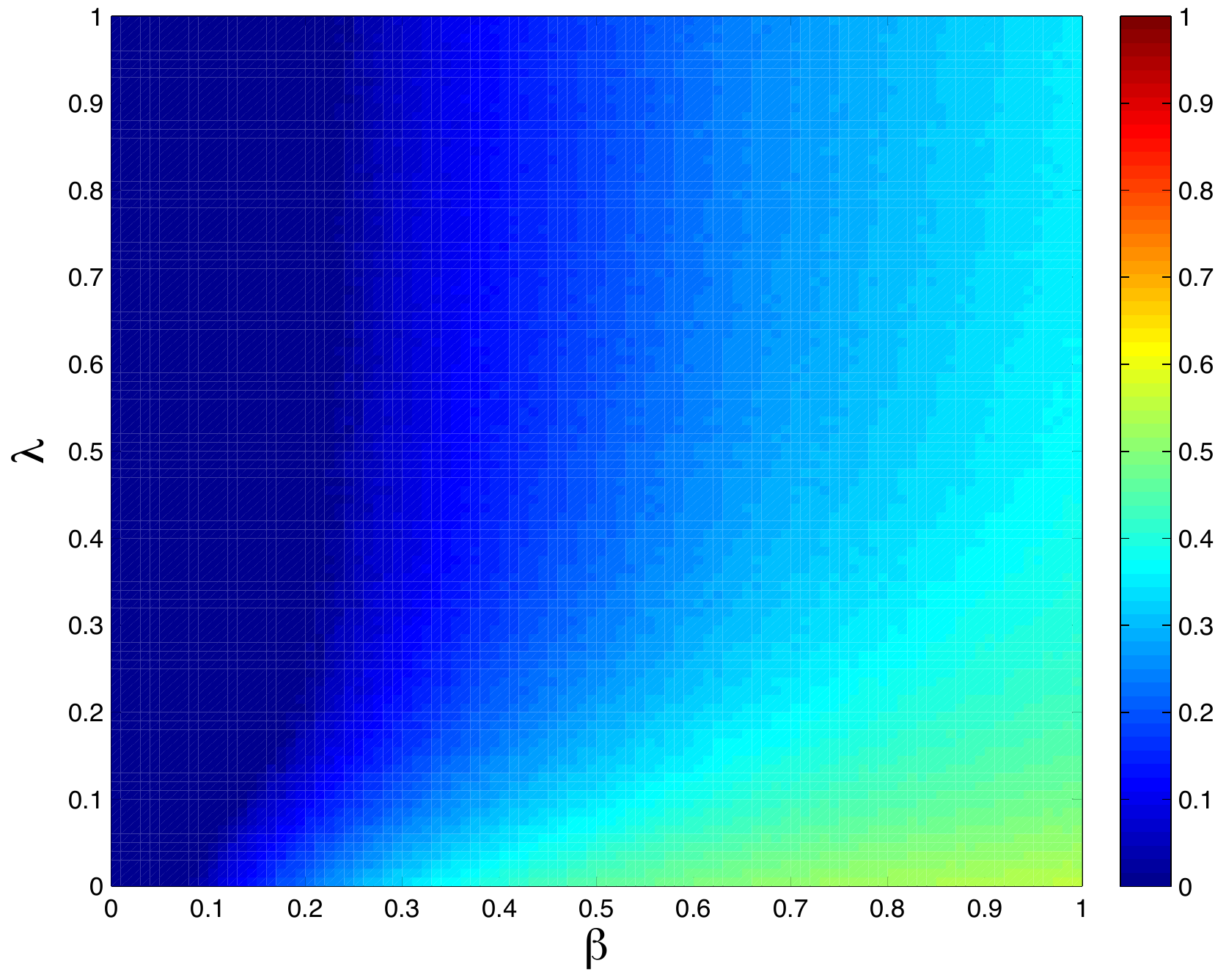}}
  \\
  MC, $\delta=0.6$, $\mu=0.4$ & MMCA, $\delta=0.6$, $\mu=0.4$ \\
  \mbox{\includegraphics*[width=0.45\textwidth,clip=0]{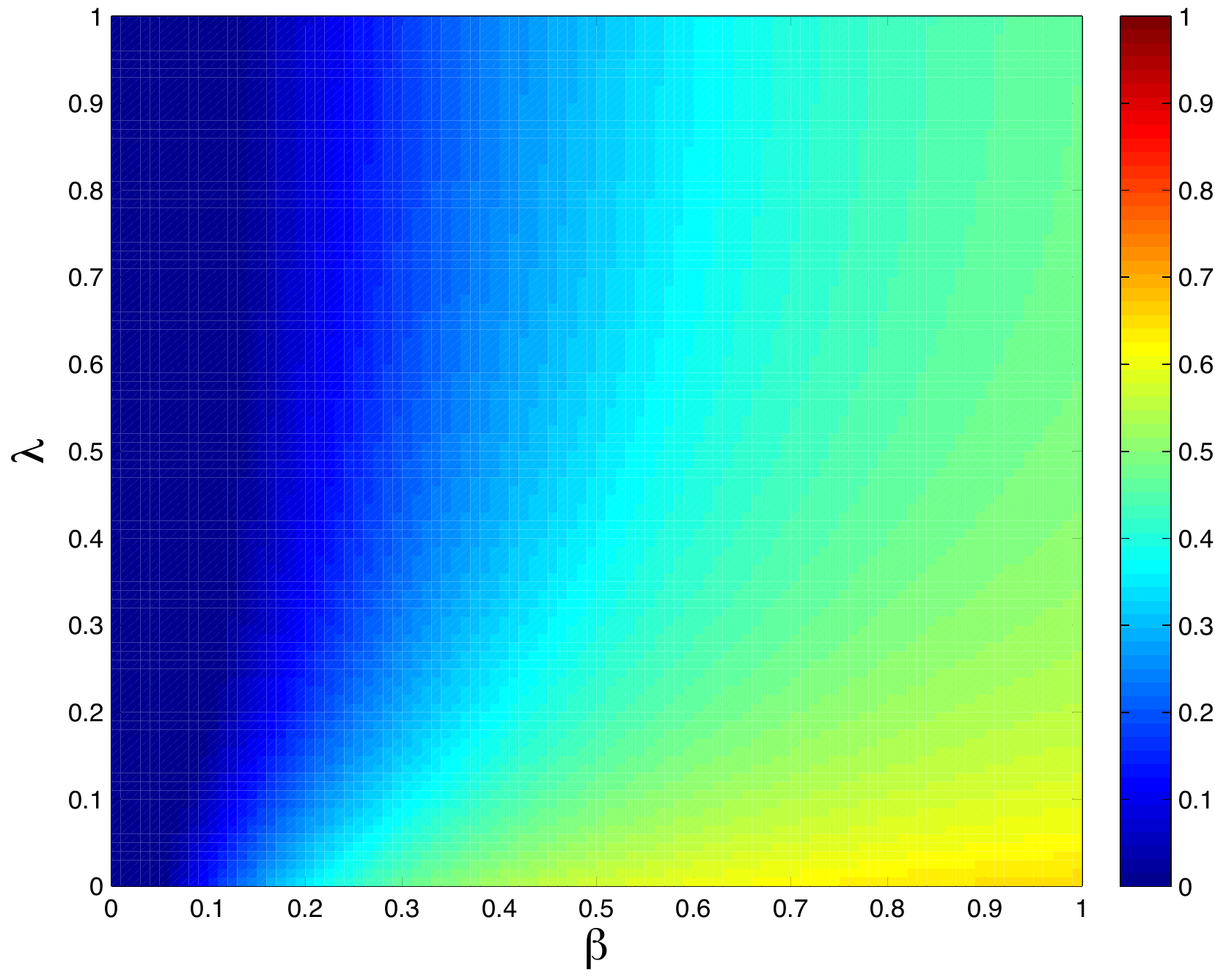}} &
  \mbox{\includegraphics*[width=0.45\textwidth,clip=0]{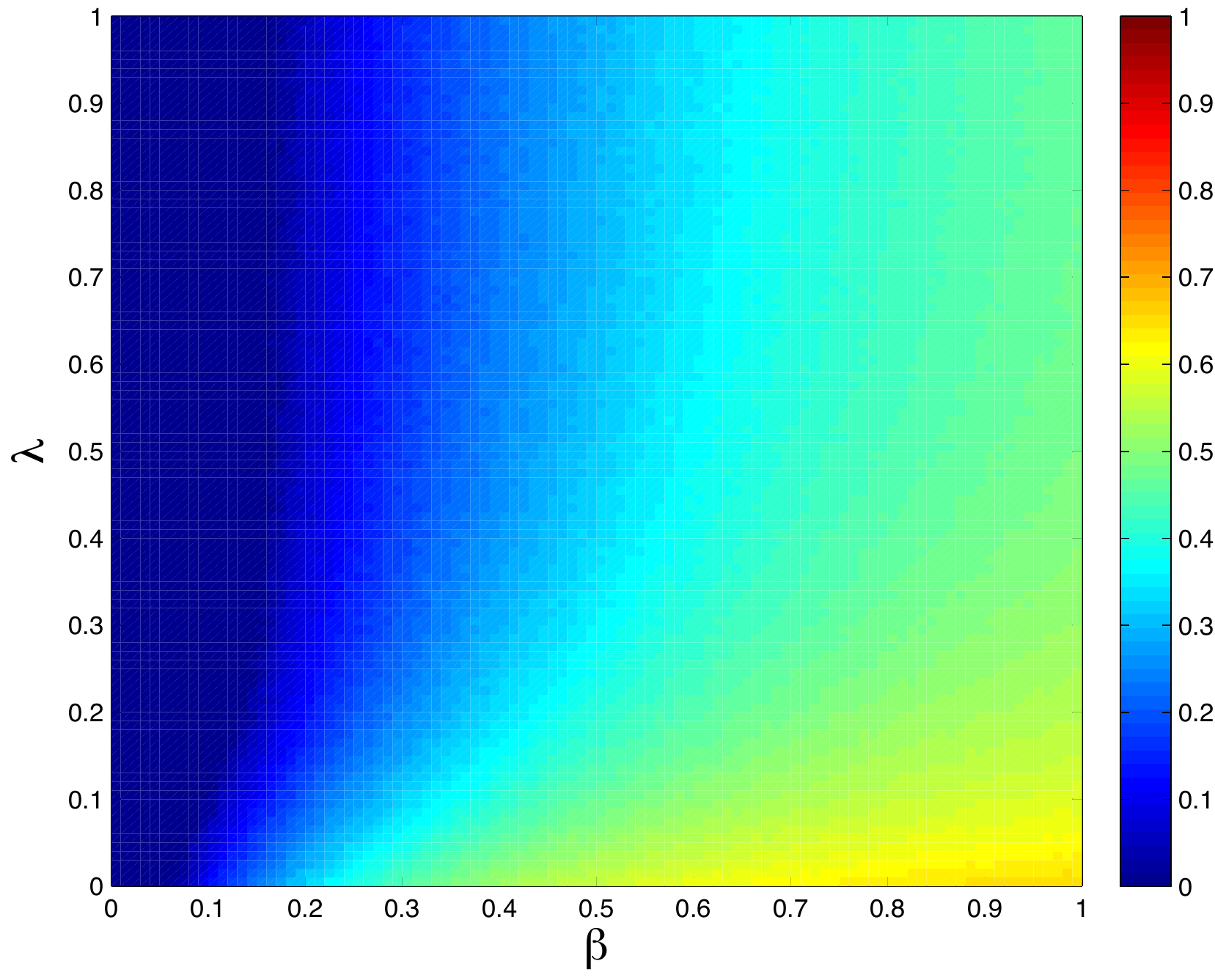}}
\end{tabular}
\caption{(color online) Comparison between Monte Carlo and MMCA for the fraction $\rho^I$ of infected individuals in the stationary state. Multiplex formed by: physical layer, a scale-free network of 1000 nodes with degree distribution $P(k)\sim k^{-2.5}$, and virtual layer, same scale-free network with 400 extra (non-overlapping) random links. Full $100\times100$ $\lambda-\beta$ phase diagram. MC values are averages over 50 simulations, and initial fraction of infected nodes is $20\%$. The relative errors between MC and MMCA are: $1.2\%$, $1.5\%$, and $1.6\%$, respectively.}
\end{figure*}

%%%%%%%%%%%%%%%%%%%%%%%%%%%%%%%%%%%%%%%%%%%%%%%%%%%%%%%%%%%%%%%%%%%%%%%
\begin{figure*}[ht]
\begin{tabular}{cc}
  MC, $\delta=0.4$, $\mu=0.6$ & MMCA, $\delta=0.4$, $\mu=0.6$ \\
  \mbox{\includegraphics*[width=0.45\textwidth,clip=0]{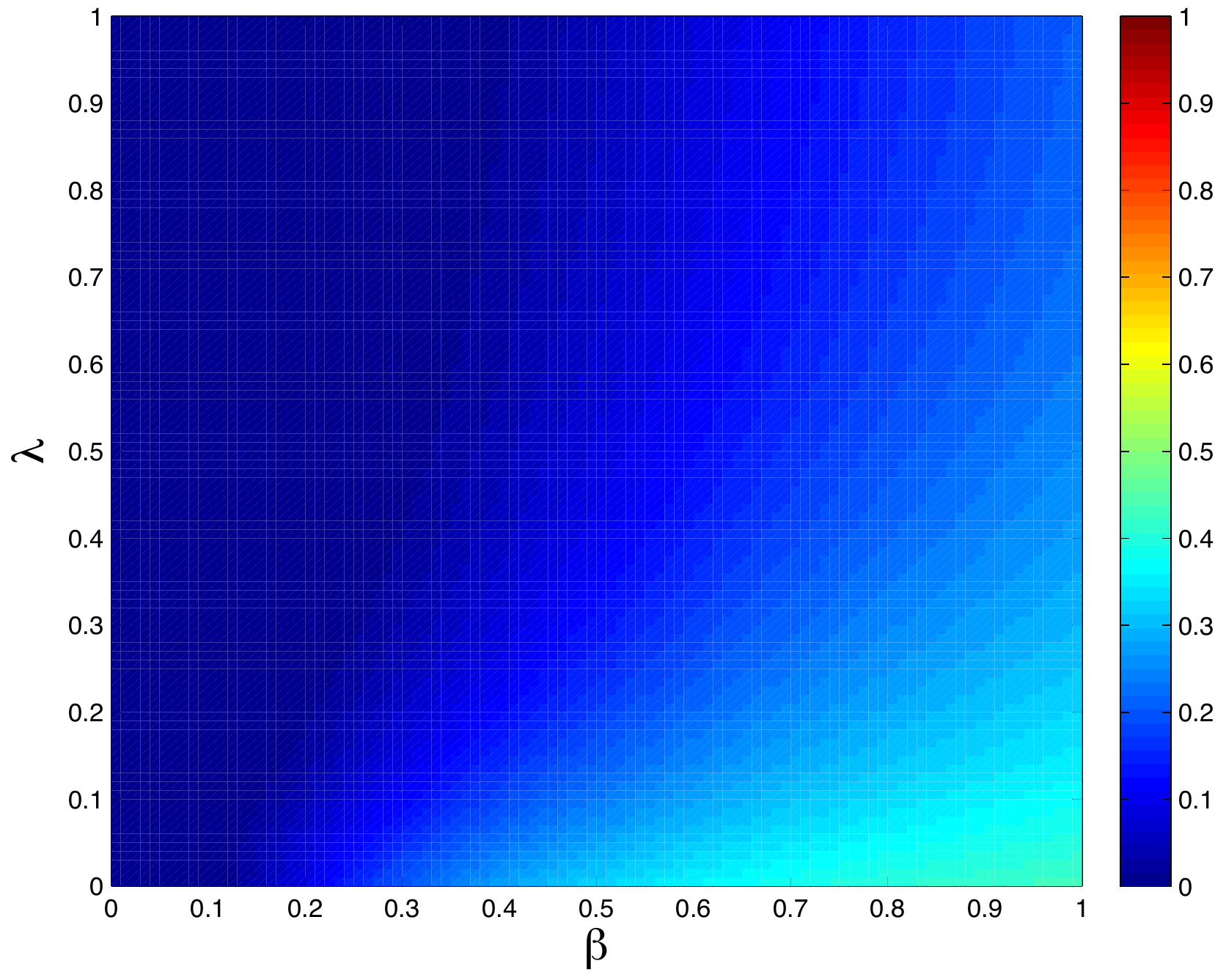}} &
  \mbox{\includegraphics*[width=0.45\textwidth,clip=0]{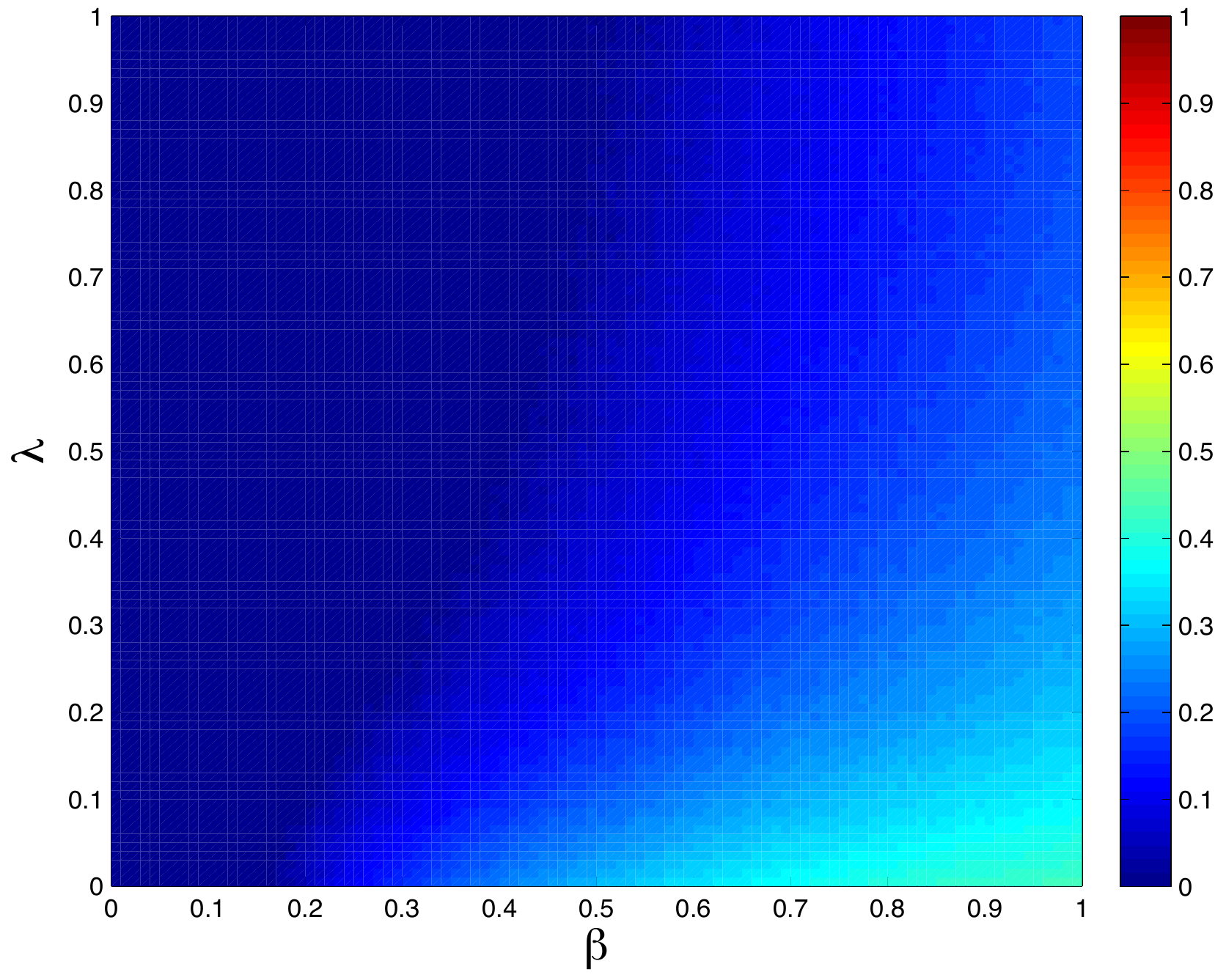}}
  \\
  MC, $\delta=0.5$, $\mu=0.5$ & MMCA, $\delta=0.5$, $\mu=0.5$ \\
  \mbox{\includegraphics*[width=0.45\textwidth,clip=0]{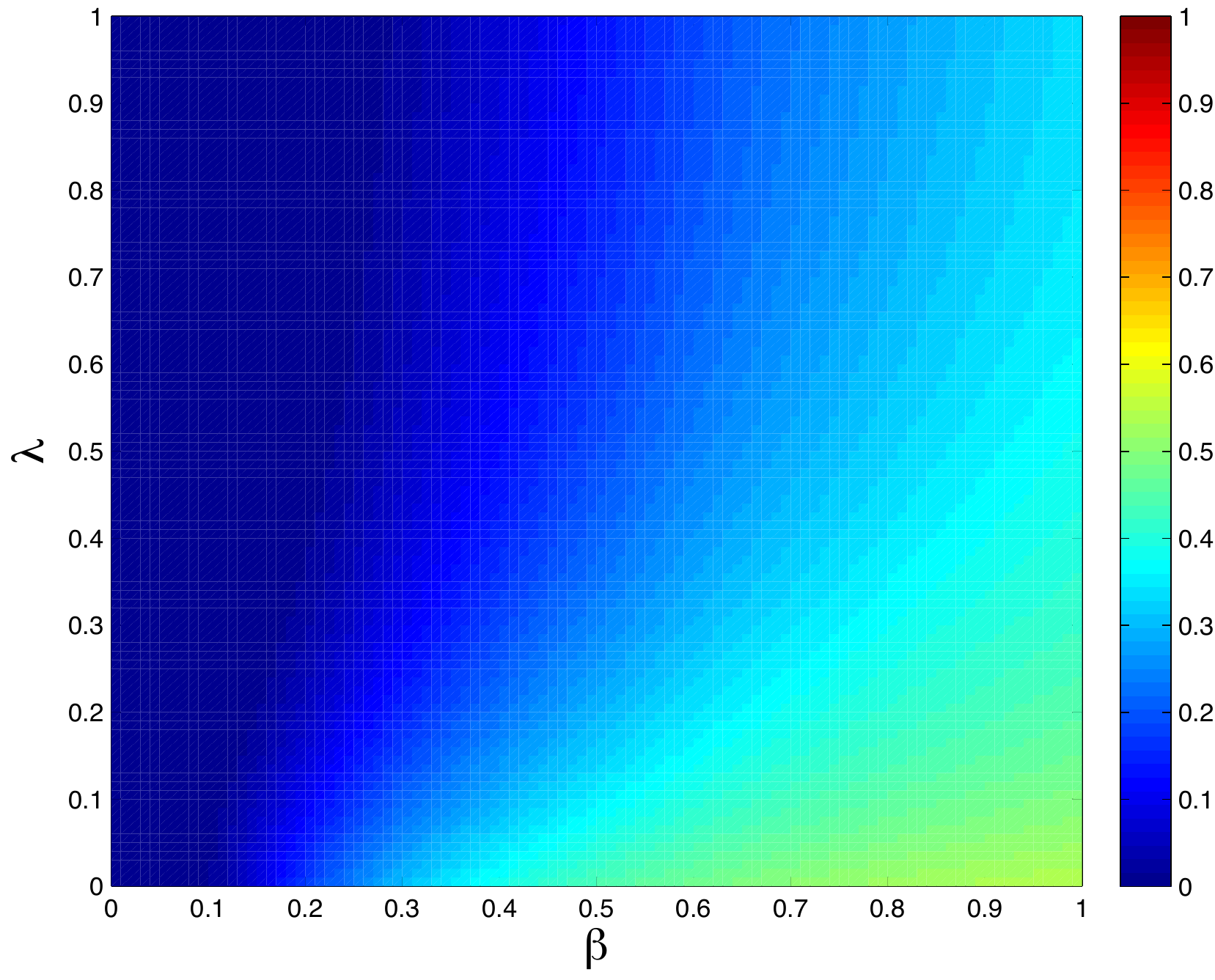}} &
  \mbox{\includegraphics*[width=0.45\textwidth,clip=0]{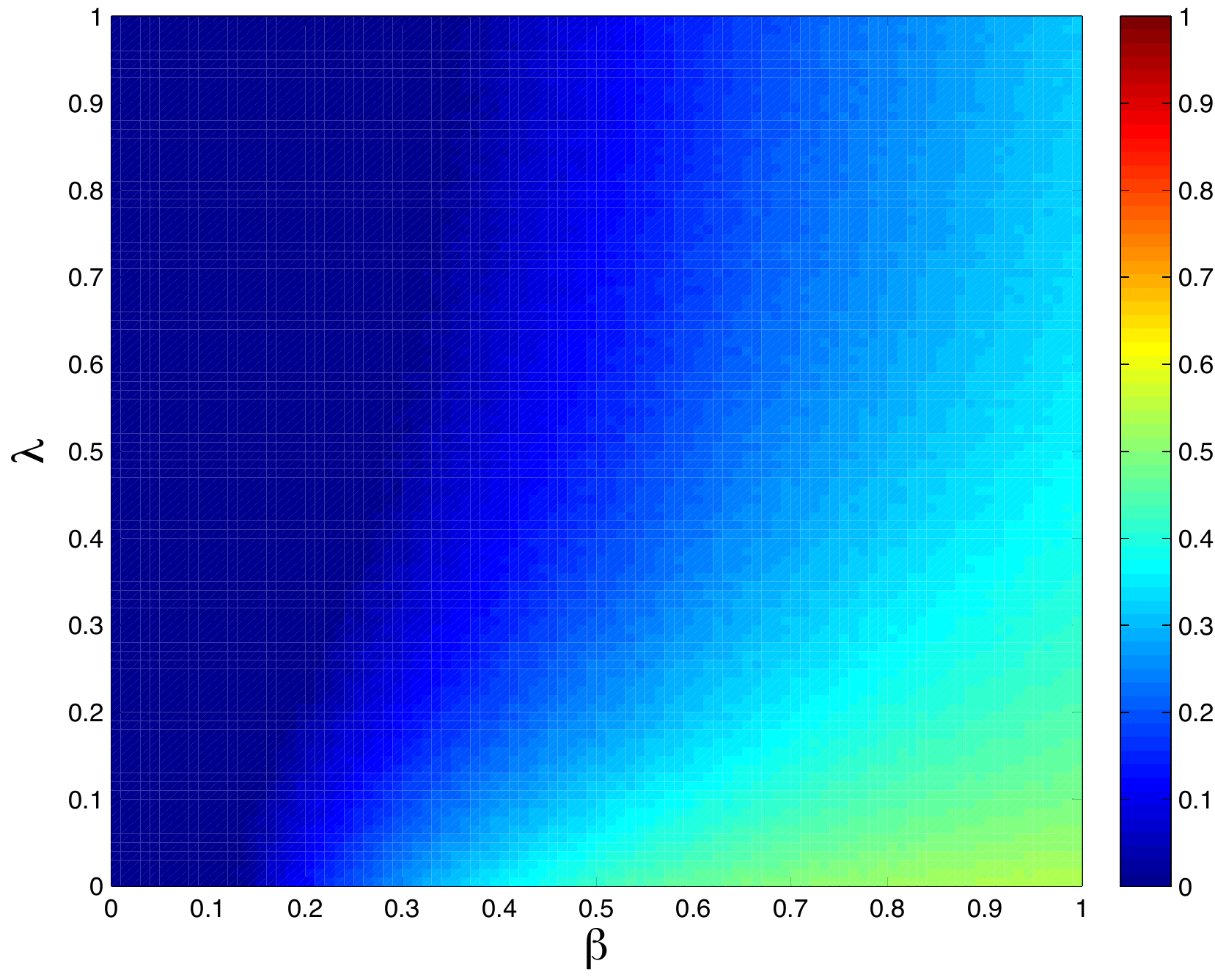}}
  \\
  MC, $\delta=0.6$, $\mu=0.4$ & MMCA, $\delta=0.6$, $\mu=0.4$ \\
  \mbox{\includegraphics*[width=0.45\textwidth,clip=0]{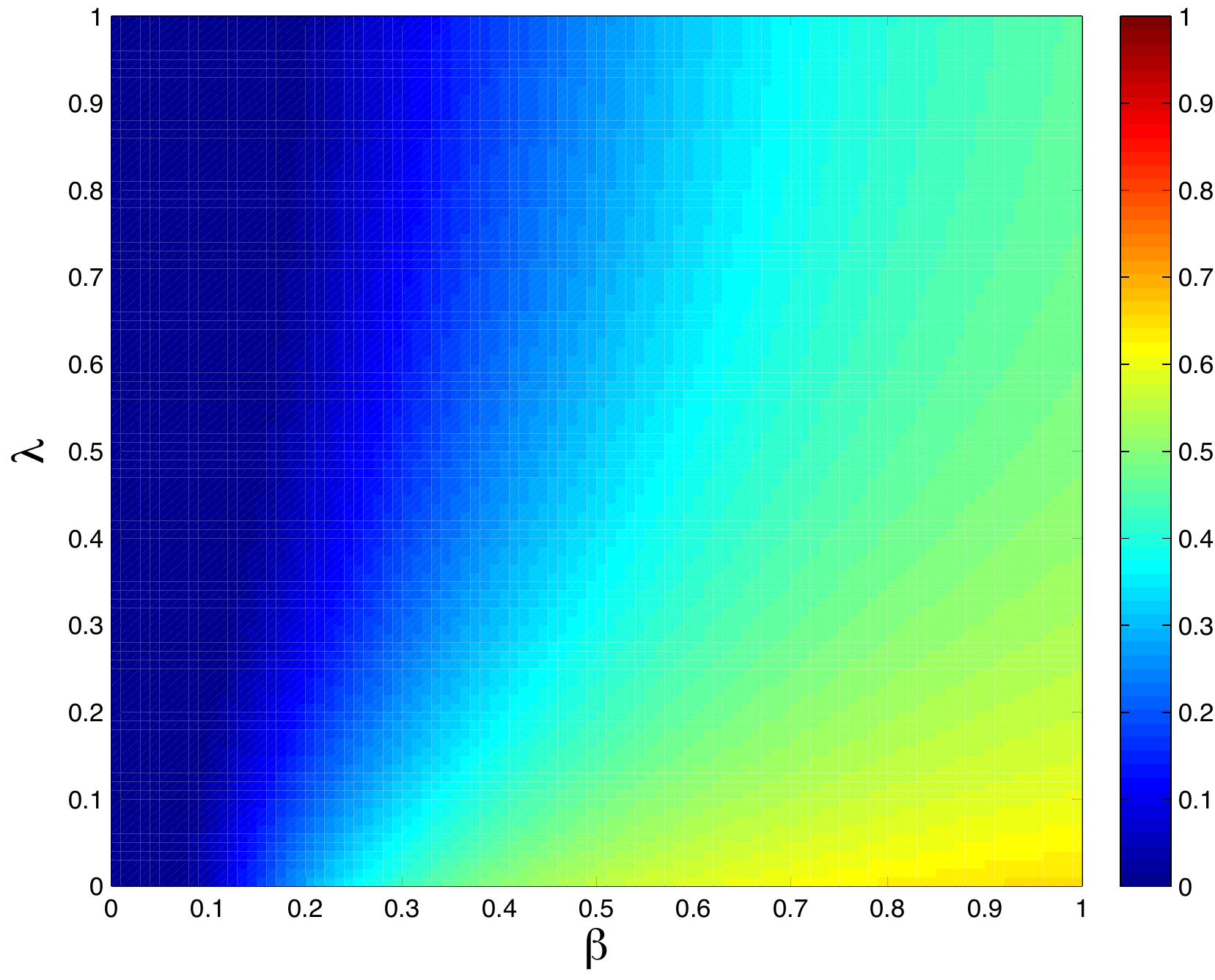}} &
  \mbox{\includegraphics*[width=0.45\textwidth,clip=0]{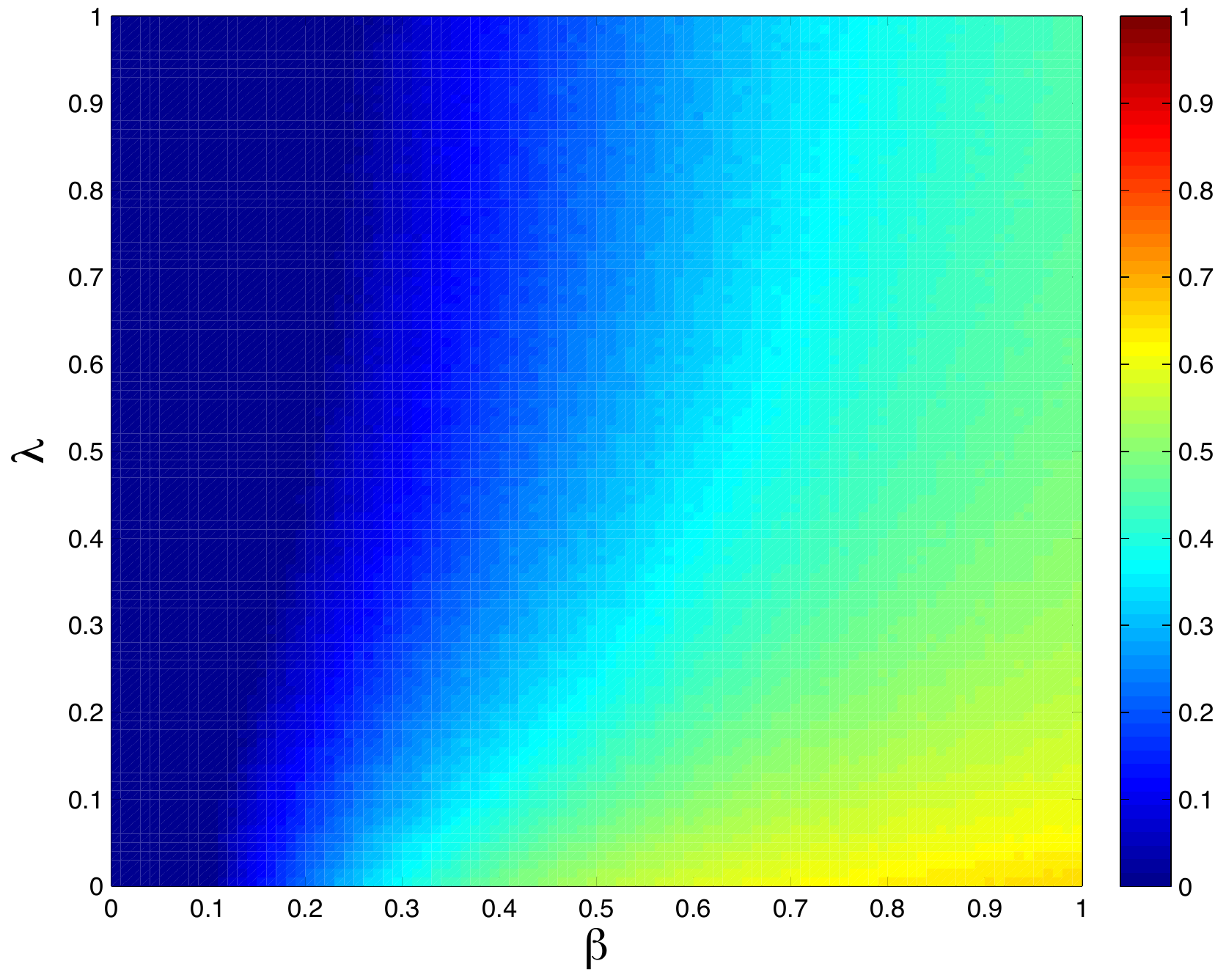}}
\end{tabular}
\caption{(color online) Comparison between Monte Carlo and MMCA for the fraction $\rho^I$ of infected individuals in the stationary state. Multiplex formed by: virtual layer, a scale-free network of 1000 nodes with degree distribution $P(k)\sim k^{-2.5}$, and physical layer, a scale-free network of 1000 nodes with degree distribution $P(k)\sim k^{-3.0}$. Full $100\times100$ $\lambda-\beta$ phase diagram. MC values are averages over 50 simulations, and initial fraction of infected nodes is $20\%$. The relative errors between MC and MMCA are: $1.9\%$, $2.3\%$, and $2.5\%$, respectively.}
\end{figure*}

\end{document}